\documentclass[floats,aps,twocolumn,showpacs]{revtex4}
\usepackage{dcolumn}
\usepackage{epsfig}
\usepackage{longtable}
\newcommand {\la} {\langle}
\newcommand {\ra} {\rangle}
\newcommand {\beq} {\begin{eqnarray}}
\newcommand {\eeqn} [1] {\label{#1} \end{eqnarray}}%
\newcommand {\eol} {\nonumber \\}
\newcommand {\ve} [1] {\mbox{\boldmath $#1$}}

\begin{document}
%
%\initfloatingfigs
%

%\tighten
\title{
Isospin symmetry in mirror $\alpha$-decays }

\author{
N.\ K.\ Timofeyuk$^{1)}$, P. Descouvemont$^{2)}$
and R.\ C.\ Johnson$^{1)}$
% \thanks is optional - remove next line if not needed
%\thanks{\emph{Present address:} Insert the address here if needed}%
}                 

\affiliation{
$^{1)}$ Department of Physics, School of Electronics and Physical Sciences, 
University of Surrey, Guildford,
Surrey GU2 7XH, UK\\
$^{2)}$ Physique Nucl\'{e}aire Th\'{e}orique et Physique 
Math\'{e}matique, CP229\\
Universit\'{e} Libre de Bruxelles, B1050 Brussels, Belgium
}

\date{\today}

\begin{abstract}

We show that a consequence of
isospin symmetry, recently discovered in mirror conjugated
one-nucleon decays, can be extended to mirror-conjugated
$\alpha$-particles decays, both virtual and real. 
For virtual $\alpha$-decays of bound mirror pairs
this symmetry manifests itself as a relation between the
Asymptotic Normalization Coefficients (ANCs)
of $\alpha$-particle overlap integrals.
This relation is given by a simple analytical
formula which  involves  
$\alpha$-particle separation energies and charges of residual nuclei.
For bound-unbound mirror pairs,  the ANC of a bound
nucleus is related to the $\alpha$-width of the mirror unbound level.
For unbound mirror pairs we get a new analytical formula that relates the
widths of mirror resonances.
We test the validity of these analytical formuli against the predictions
of a two-body potential and of
a many-body microscopic cluster model  for several mirror 
states in $^7$Li-$^7$Be, $^{11}$B-$^{11}$C and $^{19}$F-$^{19}$Ne
isotopes. We show that these analytical formulae are
valid in many cases but that some deviations can be expected for isotopes
with strongly deformed and easily excited
cores. In general, the results from microscopic model
are not very sensitive to model assumptions and can be used
to predict unknown astrophysically relevant cross sections using known
information about mirror systems.

\end{abstract}

\maketitle

\section{Introduction}

In the last few years, it has been  realised that charge symmetry of 
nucleon-nucleon (NN) interaction leads to specific relations
between the amplitudes of
mirror-conjugated  one-nucleon decays
$^A_NZ \rightarrow ^{A-1}_{N-1}Z  + n$ and
$^A_ZN \rightarrow ^{A-1}_{Z-1}N + p$ \cite{Tim03}.  In a mirror
pair of bound states
this symmetry links  Asymptotic Normalization 
Coefficients (ANCs) for mirror-conjugated overlap integrals
$\la ^A_NZ|^{A-1}_{N-1}Z \otimes n\ra$  and 
$\la ^A_ZN|^{A-1}_{Z-1}N\otimes p\ra$. In 
bound-unbound mirror states, it manifests itself as a link between the 
neutron ANC and the width of the mirror proton resonance. In both cases
this link can be represented by an  approximate simple model-independent
analytical formula that contains only nucleon binding
energies, nuclear charges and the range of the strong nucleon-core
interaction \cite{Tim03}. Comparison with microscopic cluster model calculations
\cite{Tim05a,Tim05b}
has shown that the average accuracy of this formula is about 7$\%$ for
bound mirror pairs \cite{Tim05a} and 10$\%$ for bound-unbound mirror pairs
\cite{Tim05b}.

The knowledge of the
link between mirror ANCs can be beneficial for predicting unknown
ANCs using the unformation about known mirror ANCs. The latter can be used in
nuclear astrophysics to predict or verify nucleon capture
cross sections at stellar energies.  Thus, the proton ANCs for $^8$B,
$^9$C,  
$^{12}$N and $^{27}$P have been
determined using the measured neutron ANCs for their mirror analogs $^8$Li 
\cite{Tra03}, $^9$Li \cite{Guo05}, $^{12}$B \cite{TI} and $^{27}$Mg
\cite{Guo06} respectively, and then have been used to predict 
the astrophysical $S$-factors for the corresponding 
non-resonant (p,$\gamma$) reactions on
$^7$Be, $^8$B, $^{11}$C and $^{26}$Si   at   low energies.
Also, the isospin symmetry in bound-unbound mirror pairs has been used to
predict the neutron ANC for the halo  nucleus $^{15}$C($\frac{1}{2}^+$) 
and the low-energy
cross section  for the $^{14}$C(n,$\gamma)^{15}$C($\frac{1}{2}^+$)
 reaction  using 
the measured width of the proton resonance $^{15}$F($\frac{1}{2}^+$)
\cite{Tim06a}.

In this paper, we show that similar consequences of  isospin symmetry 
are present in mirror-conjugated
$\alpha$-decays. Their knowledge  may
be used in nuclear astrophysics to predict important ($\alpha,\gamma$),
($\alpha$,N) and (N,$\alpha$) cross sections. 

In Sec.II.A we consider bound mirror pairs and
derive a simple analytical formula for the ratio of mirror ANCs squared.
As in the case of nucleon decays, the formula
 depends only on mirror  $\alpha$-particle binding energies, 
 nuclear charges and the range of the $\alpha$-core
potential. We test this formula for the
two-body model, where exact numerical solutions are available.
In Sec. II.B we make
predictions in the microscopic cluster model (MCM) for the ANCs of bound mirror
pairs $^7$Li-$^7$Be, $^{11}$B-$^{11}$C and $^{19}$F-$^{19}$Ne in which the
$\alpha$-decay threshold in the lowest. All three mirror pairs are important
for nuclear astrophysics applications. 
In Sec. III we consider bound-unbound mirror
states of the same pairs of nuclei both in a two-body model and in the MCM.
In Sec. IV we discuss isospin
 symmetry in mirror resonance states and in Sec. V we 
summarise the results obtained and draw conclusions.

%%%%%%%%%%%%%%%%%%%%%%%%%%%%%%%%%%%%%%%%%%%%%%%%%%%%%%%%%%%%%%%%%%%%%%%%%%%%%

\section{Bound mirror pairs}

\subsection{Two-body model with charge-independent $\alpha$-core strong
interaction}

We consider (1) a bound   system
$ ^{A-4}_{Z-2} (N-2) + \alpha $ and (2) its bound mirror analog  
$^{A-4}_{N-2} (Z-2)+ \alpha $ in a two-body model. We order these systems
is such a way that the binding energy $\varepsilon_1$ of the first
system is larger than the second binding energy $\varepsilon_2$. 
We denote this two cores as $X_1$ and $X_2$ and assume that
the nuclear $\alpha-X_i$ interaction $V_N$ in mirror systems
is exactly the same so that all the difference in the
wave functions $\Psi_1$ and $\Psi_2$ of these mirror systems 
is determined by different 
Coulomb interactions  $V_{C_1}$ and $V_{C_2}$. In practice, 
the  two mirror $\alpha$-particle wave functions are close to each other
both in the internal nuclear region and on the surface, where the $\alpha
-X_i$ potential strongly decreases.

The wave function $\Psi_i$, where $i$ = 1,2, satisfies the Schr\"odinger
equation 
\beq
 (T + V_N + V_{C_i} +\varepsilon_i) \Psi_i = 0
\eeqn{SchE}
with    binding energy $\varepsilon_i$. The  radial part  
$\Psi^{(i)}_l(r)$ corresponds
to the orbital momentum $l$ behaves asymptotically as
\beq
\Psi^{(i)}_l(r) \approx C^{(i)}_l W_{-\eta_i,l+1/2}(2\kappa_i r)/r .
\eeqn{as}
Here $C^{(i)}_l$ is the $\alpha$-particle ANC, $W$ is the Whittaker
function, $\kappa_i = \sqrt{2\mu \varepsilon_i }/\hbar$,
 $\mu$ is the reduced mass
for the $ \alpha + X_i$ system (we neglect the $i$ dependence
of $\mu$) and 
$\eta_i = Z_iZ_{\alpha}e^2\mu/\hbar^2\kappa_i$. 

The ANC $C^{(i)}_l$
can be represented by the integral
\beq
C^{(i)}_l = - \frac{2\mu}{\hbar^2} \int_0^{\infty} dr \, r^2 
\tilde{\phi}_l^{(i)}(r)
(V_N+V_{C_i}-\tilde{V}_{i})\Psi^{(i)}_l(r),
\eeqn{anc}
where the function $\tilde{\phi}_l^{(i)}$  is the regular
solution of the Schr\"odinger
equation with an  arbitrary  potential
$\tilde{V}_i$ 
\beq
(T_l +  \tilde{V}_{i}  +\varepsilon_i) \tilde{\phi}_l^{(i)} = 0,
\eeqn{SchEc}
with the boundary condition
\beq
\tilde{\phi}_l^{(1)}(r) \rightarrow
 \phi_l^{(1)}(r) = e^{-\frac{\pi i}{2} (l+1+\eta_1)}
F_l(i\kappa_1 r)/\kappa_1 r,
\eeqn{boundarycondition}
for  $r \rightarrow \infty$, where $F$ is the regular Coulomb function.
The only requirement on the potential $\tilde{V}_i$ is that at large
distances $r$ it should cancel the long-range Coulomb interaction potential
$V_{C_i}$ between $\alpha$ and $X_i$ in order to provide convergence
for the integral (\ref{anc}).

We exploit the freedom in choosing the $\tilde{V}_1$ to separate out
from the formula (\ref{anc}) for $C^{(2)}_l$ a term which looks as close
as possible to the corresponding formula for $C^{(1)}_l$. We
choose $\tilde{V}_1$ to be the Coulomb interaction 
$V_{C_0}^{(1)}$ between a point $\alpha$-particle and a point
core $X_1$ so that
\beq
\tilde{\phi}_l^{(1)}(r) = \phi_l^{(1)}(r) = e^{-\frac{\pi i}{2} (l+1+\eta_1)}
F_l(i\kappa_1 r)/\kappa_1 r 
\eeqn{phi}
for all $r$. We next choose $\tilde{V}_2$ so that 
$\tilde{\phi}_l^{(2)}(r)$ is proportional to
$\tilde{\phi}_l^{(1)}(r)$ for a range of values of $r  < a$ that
will be specified later. For $r > a$   the
general requirement for the  $\tilde{V}_2$ at large distances must
be satisfied, so we define
\beq
\tilde{V}_2  =    \varepsilon_1 - \varepsilon_2 +  V_{C_0}^{(1)}, & r < a
\eol
\tilde{V}_2  =   V_{C_0}^{(2)}, & r \geq a,
\eeqn{tildeV}
With this choice in Eq. (\ref{SchEc})  the function $\tilde{\phi}_l^{(2)}(r)$ 
is the regular solution of the Schr\"odinger equation
\beq
(T_l +  V_{C_0}^{(1)} +\varepsilon_1 ) \tilde{\phi}_l^{(2)}(r)= 0, & r < a
\eol
(T_l +  V_{C_0}^{(2)} +\varepsilon_2 ) \tilde{\phi}_l^{(2)}(r) = 0, & r \geq a.
\eeqn{SchEmod2}
and is therefore proportional
to $\phi_l^{(1)}(r)$ for $r < a$. Its explicit form is
\beq
\tilde{\phi}_l^{(2)}(r) = A\phi_l^{(1)}(r), & r \leq a
\eol
\tilde{\phi}_l^{(2)}(r) = \phi_l^{(2)}(r)
+ BW_{-\eta_2,l+1/2}(2\kappa_2 r)/r . 
& r \geq a
\eeqn{phimod}
The coefficients $A$ and $B$ are found from  continuity of  
$ \tilde{\phi}_l^{(2)}(r)$ and its derivative at $r = a$:
\beq
A = A_0(a) + BW_2/a\phi_l^{(1)}
%(W_2'\phi_l^{(2)}- W_2\phi_l'^{(2)})/( W_2'\phi_l^{(1)}- W_2\phi_l'^{(1)})
%_{r = a},
\eeqn{A}
where
\beq
A_0(a) = \phi_l^{(2)}(a)/\phi_l^{(1)}(a),
\eeqn{a0}
\beq
%B = a (\phi_l'^{(1)}\phi_l^{(2)} - \phi_l^{(1)} \phi_l'^{(2)})/(
%\phi_l'^{(1)}W_2 - \phi_l^{(1)} W_2')_{r = a}.
B = A^{\prime}_0(a) /(W_2/a \phi_l^{(1)})^{\prime}
\eol
\eeqn{B}
Here the notation $W_2$ for $W_{-\eta_2,l+1/2}(2\kappa_2 r)$ is 
introduced and the $^{\prime}$ denotes the differentiation
 with respect to $a$.
With these choices for the $\tilde{V}_i$ the formula (\ref{anc}) becomes
\beq
- \frac{\hbar^2}{2\mu} C^{(2)}_l =  A \int_0^a dr \, r^2 
\phi_l^{(1)}(V_N+\Delta V_{C_1})\Psi^{(2)}_l
\eol
+
\int_a^{\infty} dr \, r^2 
\tilde{\phi}_l^{(2)}
(V_N+\Delta V_{C_2})\Psi^{(2)}_l 
+ R_C(a)
%A\int_0^a dr \, r^2 
%\phi_l^{(1)}
%(V_{C_2}-V_{C_1}- \varepsilon_1 + \varepsilon_2)\Psi^{(2)}_l,
\eeqn{c2}
where
\beq
\Delta V_{C_i} = V_{C_i}-V^{(i)}_{C_0}
\eeqn{DeltaVC}
and
\beq
R_C(a) =  A
\int_0^a dr \, r^2 \phi_l^{(1)}
(V_{C_2}-V_{C_1}- \varepsilon_1 + \varepsilon_2)\Psi^{(2)}_l .
\eol
\eeqn{RC}
Introducing new functions
\beq
\Delta \Psi_{12} =  \Psi^{(2)}_l  - \Psi^{(1)}_l 
\eeqn{deltapsi12}
and
\beq
\delta \phi_{12}(r,a) = \phi_l^{(2)}(r) - A_0(a) \phi_l^{(1)}(r) ,
\eeqn{deltaphi12}
and rearranging all terms in Eq. (\ref{c2}) is such a way that
integrals from $a$ to $\infty$ do not contain 
products $\phi_l^{(1)}(r) \Psi^{(2)}_l(r)$ which increase with $r$,
we get
\beq
- \frac{\hbar^2}{2\mu} C^{(2)}_l =  A_0(a) 
\int_0^{\infty} dr \, r^2 
\phi_l^{(1)}(V_N+\Delta V_{C_1)})\Psi^{(1)}_l
\eol
+  R_C(a) + R_{\Delta\Psi} + R_{\delta \phi}(a) + R_{B}(a) + 
R_{\Delta V_C}(a),
\eol
\eeqn{c2.1}
where
the first term of the r.h.s. of the Eq. (\ref{c2.1} )
is nothing but $- \hbar^2/2\mu A_0(a) C^{(1)}_l$.

We will show that all the five remainder terms 
in Eq. (\ref{c2.1}) are small compared with either 
$- \hbar^2/2\mu A_0(a) C^{(1)}_l$ or  
$- \hbar^2/2\mu   C^{(2)}_l$ provided the  radius $a$
is chosen in a specific way.

The term $R_C(a)$  is negligible for $a < R_N$, where $R_N$ is
the radius of  the
nuclear interior, because  both the  Coulomb diffefence 
$V_{C_2}-V_{C_1}$ and the binding energy difference
$ \varepsilon_1 - \varepsilon_2$ are small compared with  
the nuclear potential $V_N$. For $a > R_N$,
$R_C(a)$ grows because the function  $\phi_l^{(1)}$ increases
faster than $\Psi^{(2)}_l$ decreases.

The contribution from $R_{\Delta \Psi}$, where
\beq
R_{\Delta \Psi} = 
\int_0^{\infty} dr \, r^2 
\phi_l^{(2)}(V_N+\Delta V_{C_1})\Delta\Psi_{12},
\eeqn{RDeltaPsi}
does not depend on $a$
and is determined by the difference between the
functions $\Psi^{(2)}_l$ and $\Psi^{(1)}_l$ 
in the region that gives the most contribution to the integral
in the r.h.s. of Eq. (\ref{RDeltaPsi}). %$R_{\Delta \Psi}$. 
In the cases
considered below, this difference is about 2$\%$.

The term $R_{\delta \phi} (a)$ defined as
\beq
R_{\delta \phi} (a) = 
\int_a^{\infty} dr \, r^2 
\delta \phi_{12}(r,a)V_N \Psi^{(1)}_l
\eol
-\int_0^a dr \, r^2 
\delta \phi_{12}(r,a)V_N \Delta \Psi_{12},
\eeqn{Rdeltaphi}
contains the function $\delta \phi_{12}(r,a)$ which is equal to zero
at $r=a$. Therefore, if $a$ is   at   a point where
$V_N \Psi^{(1)}_l$ reaches its maximum   and is a decreasing
function at $r > a$ then the contribution from $R_{\delta \phi} (a)$
will be small.% due to dramatic reduction of the
%integrand in the r.h.s. of Eq. (\ref{Rdeltaphi}). 
This point
can be chosen to be the nuclear radius $R_N$, which for
$\alpha +X$ system is about (1.1-1.3)(4$^{1/3}$+$X^{1/3}$). 
If at the same time  
$ \phi_l^{(2)}(r)/\phi_l^{(1)}(r)$  varies slowly  with $r$ around $a$
then $\phi_{12}(r,a) \approx 0$
which guarantees that $R_{\delta \phi} (a)$ is negligible.
However, $R_{\delta \phi} (a)$   increases if $a < R_N$
and   $ \phi_l^{(2)}/\phi_l^{(1)}$ at $r = R_N$ differs
from $A_0(a)$. On the other hand, $R_{\delta \phi} (a)$
is very small for $a > R_N$.
 
The next term, 
\beq
R_B(a) =B 
\int_a^{\infty} dr \, r\, W_2 
(V_N+\Delta V_{C_2})\Psi^{(2)}_l
\eol
+ B \frac{W_2}{a\phi_l^{(1)}} 
\int_0^a dr \, r^2 
\phi_l^{(1)}(V_N+\Delta V_{C_1})\Psi^{(2)}_l,
\eeqn{RB}
depends on $B$. The $B$ is  zero at two points, at  $a = 0$
and at  
$a = a_m$  where the function $A_0(a)$ reaches its maximum (or
in other words $A^{ \prime} _0 (a_m) = 0$).
At all other points the
contribution from $R_B(a)$ depends on how large is
$BW_2/a\phi^{(1)}_l$ with respect to $A_0(a)$. We show in
Appendix that
\beq
\frac{BW_2 }{a\phi^{(1)}_lA_0(a)} = \frac{p_2(a)-p_1(a)}{p_2(a)+p_1(a)},
\eeqn{baratio}
where
\beq
p_i(a) = \sqrt{\frac{2\eta_i\kappa_i}{r} + \frac{l(l+1)}{r^2} +
\kappa_i^2}.
\eeqn{pi}
For mirror $\alpha$ states $p_2(a)$ does not differ much from
$p_1(a)$, especially near $a \approx R_N$. 
Thus $BW_2/a\phi^{(1)}_l << A_0(a)$  and, therefore, $R_B(R_N)$ will
be small compared with  $- \hbar^2/2\mu A_0(a) C^{(1)}_l$.

The last term, 
\beq
R_{\Delta V_C}(a) = 
\int_a^{\infty} dr \, r^2 
(\phi_l^{(2)}\Delta V_{C_2}
-A_0(a)\phi_l^{(1)}
\Delta V_{C_1})  \Psi^{(1)}_l
\eol
-\int_0^a dr \, r^2 
(\phi_l^{(2)}\Delta V_{C_2}
-A_0(a)\phi_l^{(1)}
\Delta V_{C_1})  \Delta \Psi_{12}.
\eol
\eeqn{RDeltaVC}
is zero for all $a$ greater than the radius of the $\alpha$-core
Coulomb interaction $R_c$
and is small for $a < R_c$ if $\Delta V_{C_i} \ll V_N $. For all cases
considered below, this condition is satisfied.

Thus, if $\Psi^{(1)}_l \approx \Psi^{(2)}_l$ is a good approximation
and if $a$ is chosen near $R_N$ then the contributions from
all the remainder terms  $R_i(a)$ are very small and Eq.(\ref{c2.1})
reduces to
\beq
 \frac{\hbar^2}{2\mu} C^{(2)}_l =  A_0(a) 
 \frac{\hbar^2}{2\mu} C^{(1)}_l.
\eol
\eeqn{c2.2}
Then the ratio ${\cal R}$ 
\beq
{\cal R} =\left(C^{(2)}_l/C^{(1)}_l \right)^2
\eeqn{ratioC}
of the mirror squared ANCs can be approximated by the model-independent
analytical expression
\beq
{\cal R} \approx {\cal R}_0 = A_0^2(R_N) = 
\left| \frac {\kappa_1F_l(i\kappa_2R_N)}{\kappa_2
F_l(i\kappa_1R_N)}\right|^2.
\eeqn{rna}
The accuracy of this approximation depends on how rapidly
$A_0(R_N)$ changes over  the region of uncertainty of $R_N$. In all
cases considered below this function varies slowly around $R_N$
(see the insets in Fig.1 where $A_0(a)/A_0(a_m)$ is plotted).

The approximation (\ref{rna}) is similar to  the formula,
\beq
%{\cal R}_0 =
\left( \frac{C_p}{C_n}\right)^2 \approx\left(
\frac{F_l(i\kappa_pR_N)}{\kappa_pR_N\,j_l(i\kappa_nR_N)}
\right)^2,
\eeqn{rn}
obtained in Ref. \cite{Tim03}
for  ANCs $C_p$ and $C_n$ of mirror  proton and neutron virtual decays
respectively.  In principle,
Eq. (\ref{rna}) could be obtained from (\ref{rn}) 
by replacing the spherical Bessel function $j_l(i\kappa_n r)$
by $F_l(i\kappa_1R_N)/\kappa_1 R_N$. However, Eq. (\ref{rn}) has been 
obtained  in Ref. \cite{Tim03} starting from different assumptions. Namely,
it was explicitely assumed that the main contribution to the ANC
comes only from internal nuclear region, $r \leq R_N$,
that the Coulomb interactions
inside the nuclear region can be replaced by constants and that the
difference between these  constants is equal to the difference in 
proton and neutron binding energies. Our exact two-body calculations
have shown that the accuracy of these assumptions is much worse than
the accuracy of the formula (\ref{rna}) itself. In particular, all 
$\alpha$-particle wave functions have nodes because of the Pauli principle,
which causes cancellations between some contributions to the ANC from
the internal region so that the contributions from the surface become 
important. For large orbital momentum $l$ the  
surface region, in which the nuclear potential decreases, is even more
important.
We illustrate this in the insets of Fig.1 
by plotting some examples of $C_2(a)/C_2$,
where the ANC $C_2(a)$ has been calculated neglecting the
contributions from $r > a$ in Eq. (\ref{anc}). Quite often the
$r \leq R_N$ region  gives only  half  the contribution to the ANC.
The derivation of Eq. (\ref{rna}) in the present paper  is quite general
and it suggests that Eq. (\ref{rna}) should be valid even 
when the contribution
from $r \leq R_N$ is small. Also, this equation should be valid for
all shapes of nuclear potentials, even with unphysically diffused edges,
and does not depend on the exact functional form of the Coulomb potential
in the internal region. The only criteria of its applicability is
the similarity of the wave functions of mirror nuclei.

\begin{figure}[t]
%\centerline{\psfig{figure=../../alpha.fig1.eps,width=0.40\textwidth} }
\centerline{\psfig{figure=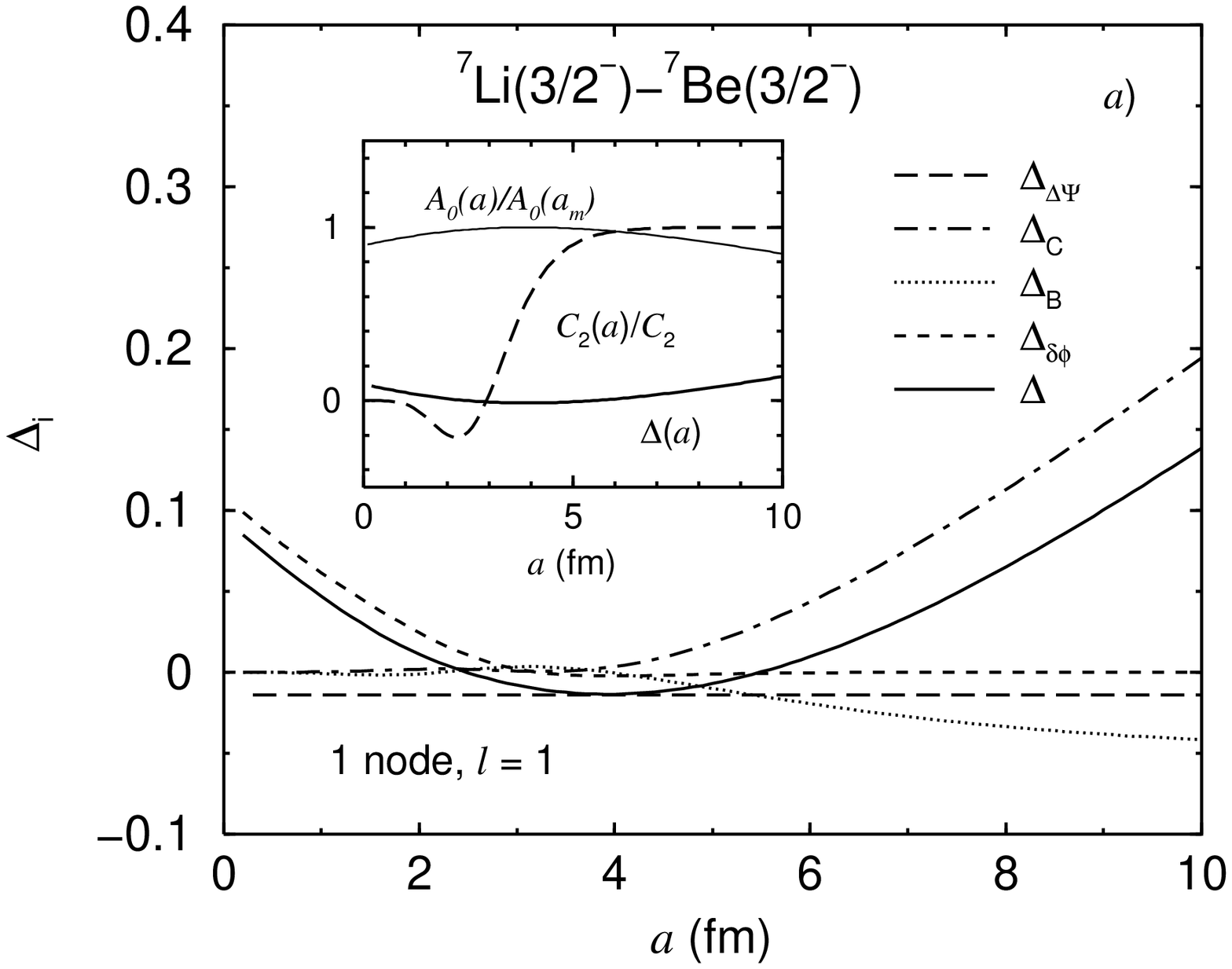,width=0.40\textwidth} }
%{\psfig{figure=../../alpha.fig2.eps,width=0.40\textwidth} }
{\psfig{figure=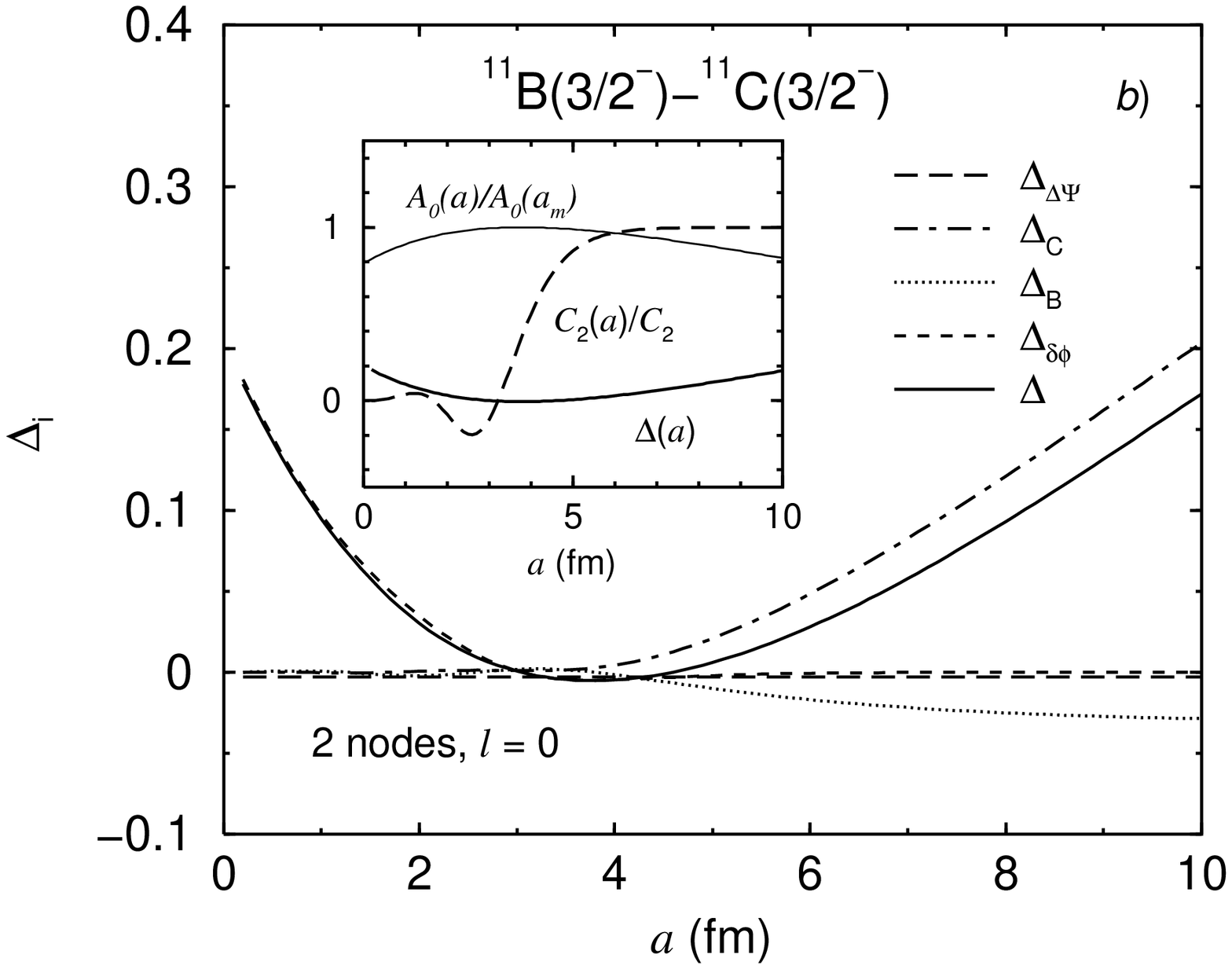,width=0.40\textwidth} }
%{\psfig{figure=../../alpha.fig3.eps,width=0.40\textwidth} }
{\psfig{figure=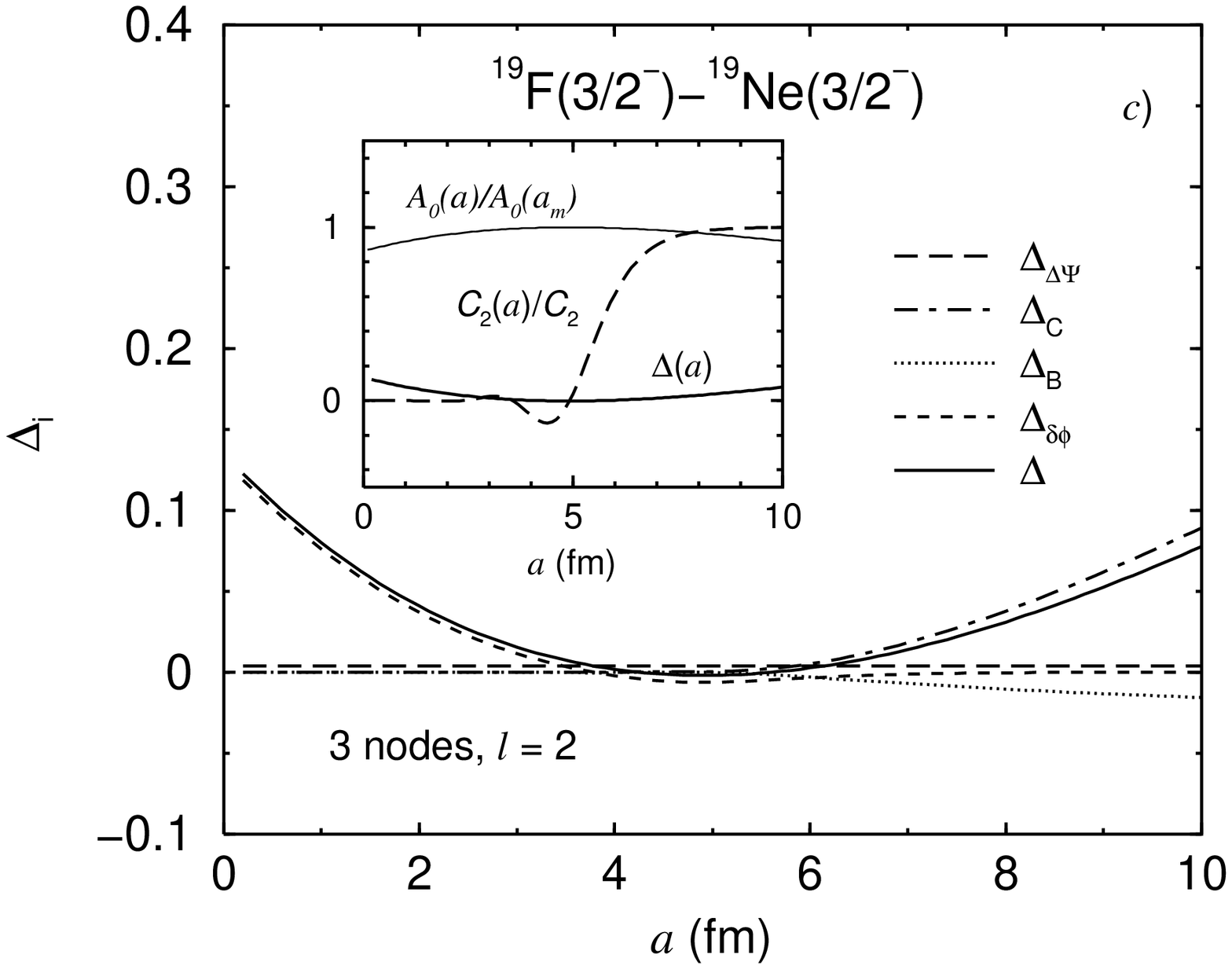,width=0.40\textwidth} }
\caption{ The deviations $\Delta_i$ and $\Delta = \sum_i \Delta_i$ 
as a function
of matching radius $a$  for the $\frac{3}{2}^-$ states in
mirror pairs $^7$Li-$^7$Be ($a$),
$^{11}$B-$^{11}$C ($b$) and
$^{19}$F-$^{19}$Ne ($c$). Also shown in insets are the ratios
$A_0(a)/A_0(a_m)$ and $C_2(a)/C_2$.
}
\end{figure}

\begin{figure*}[t]
\centerline{\psfig{figure=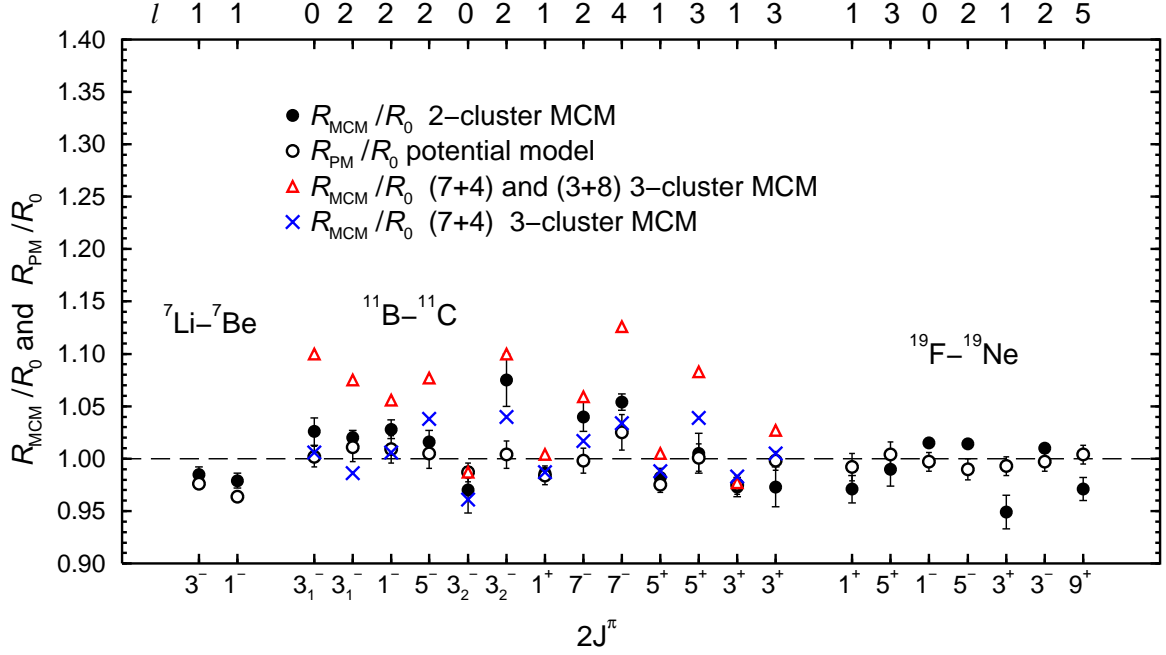,width=0.86\textwidth} }
\caption{ Ratio of the potential model estimate ${\cal R}_{PM}$ to the
analytical estimate ${\cal R}_0$ (open circles) and ratio of the
MCM predictions  ${\cal R}_{MCM}$ to the analytical estimate ${\cal R}_0$
calculated in the two-cluster (filled circles)
and three-cluster microscopic cluster model in which both
the $^7$Li+$\alpha$  ($^7$Be+$\alpha$)
and t+$^8$Be ($^3$He+$^8$Be)  (triangles) or only
the $^7$Li+$\alpha$  ($^7$Be+$\alpha$)
 partitions (crosses) have been taken into account.
}
\end{figure*}

In Fig.1 we   show  the deviations $\Delta_i$  from 
$C_l^{(2)}$ defined as
\beq
\Delta_i = -\frac{2\mu}{\hbar^2} R_i(a)/C_l^{(2)}
\eeqn{deltai}
where $i$ = $
C$, ${\Delta\Psi}$, ${\delta \phi}$, ${B}$ and
$\Delta V_C$, together with the total deviation  $\Delta = 
\sum_i \Delta_i$ for three mirror pairs, 
$^7$Li(=$\alpha$+t) - $^7$Be(=$\alpha$+$^3He$),
$^{11}$B(=$\alpha$+$^7$Li) - $^{11}$C(=$\alpha$+$^7$Be) and
$^{19}$F(=$\alpha$+$^{15}$N) -  $^{19}$Ne(=$\alpha$+$^{15}$O).
The calculations have been done using a Woods-Saxon potential with
a diffuseness of 0.65 fm, the radius and the depth of which has been
adjusted to fit the $\alpha$-particle energies in mirror systems.
The total spin-parity in all three cases is $\frac{3}{2}^-$ (the second
$\frac{3}{2}^-$ state was considered for $^{11}$B-$^{11}$C to enhance
the difference in the mirror wave functions) but the
orbital momenta $l$ and the number of nodes are different. 
The ratio  $A_0(a)/A_0(a_m)$, shown in insets of Fig.1,
does not change much near $R_N$.
The total deviation $\Delta$ is minimal at $r = R_N$ and
is determined mainly by $\Delta_{\delta \phi}$ 
when $r < R_N$ and by $\Delta_C+\Delta_B$
at $r > R_N$ with  $\Delta_C$ significantly larger than $\Delta_B$.
The contribution from $\Delta_{\Delta_{V_C}}$ 
is too small to be shown in these figures.

We have performed   exact two-body calculations for other states
of the  mirror pairs
$^7$Li-$^7$Be, $^{11}$B-$^{11}$C and $^{19}$F-$^{19}$Ne  using Woods-Saxon
potentials with diffusseness varying from 0.35 to 0.95 fm. 
%and the radius and the depths of which have been ajdusted to
%reproduce simlutaneously the $\alpha$-particle separation energies
%in these isobars. 
The sensitivity of the ratio ${\cal R}$ to the 
potential choice was less than 2$\%$. 
Both the exact ratios ${\cal R}_{PM}$
and the analytical approximations ${\cal R}_0$ 
 are given in Table I. Since
in all cases   $a_m$ was  very  close to  
$R_N$ and $A_0(a)$ changed very slowly around $R_N$, the ${\cal R}_0$ values
from Table II were calculated at $R_N = a_m$.
The ratio ${\cal R}_{PM}/{\cal R}_0$ is also plotted in Fig.2.
One can see that  ${\cal R}_{PM}$ and ${\cal R}_0$ agree  on average within
2$\%$ or less. For $^7$Li-$^7$Be this agreement is slightly worse,
about 3-4$\%$, which can be explained by the larger difference
in internal wave functions due to the smaller Coulomb interaction.

%%%%%%%%%%%%%%%%%%%%%%%%%%%%%%%%%%%%%%%%%%%%%%%%%%%%%%%

\begin{table*}[t]
\caption{ 
Microscopic calculations for
${\cal R}_{MCM}$, analytical estimate ${\cal R}_0$ 
and the potential model estimate ${\cal R}_{PM}$, 
for the mirror pairs from   the first column with the spin-parity
$J^{\pi}$ and the orbital momentum $l$ of the $\alpha$ particle. 
Also shown are
the ratios ${\cal R}_{b_{\alpha}}^{MCM} = (b_{\alpha}(2)/b_{\alpha}(1))^2$, 
where $b_{\alpha}(i) = C_{\alpha}(i)/\sqrt{S_{\alpha}(i)}$
is the normalized  ANC for the nucleus $i$, and $S_{\alpha}$
is the spectroscopic factor. The significance of these ratios is discussed
in the text. For ${\cal R}_{MCM}$ and ${\cal R}_{b_{\alpha}}^{MCM}$,
average values and range of variations
 between  calculations with V2 and MN potentials and
two different oscillator radii are
presented.  ${\cal R}_{PM}$ is averaged over the choice of different
parameters of the Woods-Saxon potentials and shown together with
the range of its variation.
 } 
\begin {center}
\begin{tabular}{ p{3 cm} p{1 cm} p{1 cm} p{3.2 cm} p{2.4 cm} p{3.2 cm}
 p{3.2 cm}}
\hline 
\\ 
Mirror pair  &  $J^{\pi}$ & $l$ %& $\epsilon_p$ & $\epsilon_n$ 
& ${\cal R}_{MCM}$  & ${\cal R}_0$ &
${\cal R}_{PM}$
 &  
 ${\cal R}_{b_{\alpha}}^{MCM}$  \\
 \hline
$^{7}{\rm Li}  - ^{7}{\rm Be}$ & $\frac{3}{2}^-$ &
1 &   1.35 $\pm$ 0.01 & 1.37 &  1.34 $\pm$ 0.01 &
1.37 $\pm$ 0.01  \\
  & $\frac{1}{2}^-$ &
1 &   1.43 $\pm$ 0.01 & 1.47 &  1.41 $\pm$ 0.01 &
1.45 $\pm$ 0.01  \\
 \\
$^{11}{\rm B}  - ^{11}{\rm C}$ & $\frac{3}{2}^-_1$ &
0 &   1.60 $\pm$ 0.02 & 1.56 &  1.57 $\pm$ 0.02 &
1.55 $\pm$ 0.01  \\
two-cluster MCM & & 2 &   1.50 $\pm$ 0.01 & 1.46 &  1.49 $\pm$ 0.02  &
1.51 $\pm$ 0.02  \\
& $\frac{1}{2}^-$ &
2 &   1.65 $\pm$ 0.02 & 1.60 &  1.61  $\pm$ 0.02  &
1.64 $\pm$ 0.02  \\
 & $\frac{5}{2}^-$ &
2 &   1.85 $\pm$ 0.02 & 1.82 &  1.83  $\pm$ 0.02  &
1.83 $\pm$ 0.02  \\
& $\frac{3}{2}^-_2$ & 0 &   2.23 $\pm$ 0.05 & 2.30 &  2.27 $\pm$ 0.02 &
2.27 $\pm$ 0.02  \\
 & & 2 &   2.16 $\pm$ 0.05 & 2.01 &  2.02 $\pm$ 0.03 &
2.06 $\pm$ 0.02  \\
 & $\frac{1}{2}^+$ &
1 &   4.55 $\pm$ 0.01  & 4.61 &  4.54 $\pm$ 0.04  &
4.54 $\pm$ 0.02  \\
& $\frac{7}{2}^-$ & 2 &   4.38 $\pm$ 0.06  & 4.20 &  4.19 $\pm$ 0.05 &
  4.24 $\pm$ 0.02 \\
 & & 4 &   2.51 $\pm$ 0.02   & 2.38 &  2.44 $\pm$ 0.04 &
 2.48 $\pm$ 0.01  \\ 
 & $\frac{5}{2}^+$ &
1 &   13.29 $\pm$ 0.12 & 13.53 &  13.19 $\pm$ 0.10 &
13.2 $\pm$ 0.1  \\
 & & 3 &   7.79 $\pm$ 0.15 & 7.75 &  7.76 $\pm$ 0.10 &
7.56 $\pm$ 0.04  \\
 & $\frac{3}{2}^+$ &
1 &  $ (1.68 \pm 0.02)  \times 10^{12}$ & 
$1.72  \times 10^{12}$ &  $(1.68 \pm 0.02)  \times 10^{12}$ &
$(1.66 \pm 0.02)  \times 10^{12}$ \\
 & & 3 &   $(3.59 \pm 0.07)  \times 10^{11}$ & 
 $3.69\times 10^{11}$ &  $(3.68 \pm 0.03)  \times 10^{11}$ &
$(3.55 \pm 0.05)  \times 10^{11}$  \\
\\
$^{11}{\rm B}  - ^{11}{\rm C}$ & $\frac{3}{2}^-_1$ &
0 &   1.71 & 1.56 &  1.56  $\pm$ 0.02 & 1.66  \\
 three-cluster MCM & & 2 &   1.58 & 1.47 &  1.49 $\pm$ 0.02  & 1.56  \\
& $\frac{1}{2}^-$ &
2 &   1.69 & 1.60 &  1.61  $\pm$ 0.02  & 1.66  \\
 & $\frac{5}{2}^-$ &
2 &   1.96 & 1.82 &  1.83  $\pm$ 0.02  &
1.91  \\
& $\frac{3}{2}^-_2$ & 0 &   2.27 & 2.30 &  2.27 $\pm$ 0.02 &
2.31  \\
 & & 2 &   2.21 & 2.01 &  2.02 $\pm$ 0.03 & 2.09  \\
 & $\frac{1}{2}^+$ &
1 &   4.63 & 4.61 &  4.54 $\pm$ 0.04  & 4.61  \\
& $\frac{7}{2}^-$ & 2 & 4.45   & 4.20 &  4.19 $\pm$ 0.05 &
4.24   \\
 & & 4 &  2.68   & 2.38 &  2.44 $\pm$ 0.04 & 2.64
   \\ 
 & $\frac{5}{2}^+$ &
1 &   13.60 & 13.53 &  13.19 $\pm$ 0.10 & 13.46  \\
 & & 3 &   8.39 & 7.75 &  7.76 $\pm$ 0.10 & 7.76  \\
 & $\frac{3}{2}^+$ &
1 &   $ 1.68\times 10^{12}$ & 
$1.72 \times 10^{12}$ &  $(1.68\pm 0.02)  \times 10^{12}$ &
$1.70 \times 10^{12}$ \\
 & & 3 &   $ 3.79\times 10^{11}$ & 
 $3.69\times 10^{11}$ &  $(3.68 \pm 0.03)  \times 10^{11}$ &
 $3.69\times 10^{11}$  \\
 \\
$^{19}{\rm F}  - ^{19}{\rm Ne}$ & $\frac{1}{2}^+$ &
1 &   4.12 $\pm$ 0.06 & 4.24 &  4.21 $\pm$ 0.06 &
4.17 $\pm$ 0.04  \\
 & $\frac{5}{2}^+$ &
3 &   4.23 $\pm$ 0.07 & 4.27 &  4.29 $\pm$ 0.04  &
4.26 $\pm$ 0.07  \\
             & $\frac{1}{2}^-$ &
0 &   4.70 $\pm$ 0.01  & 4.63 &  4.61  $\pm$ 0.04  &
4.66 $\pm$ 0.01   \\
 & $\frac{5}{2}^-$ &
2 &   9.58 $\pm$ 0.04 & 9.44 &  9.43 $\pm$ 0.09  &
9.53 $\pm$ 0.02  \\
 & $\frac{3}{2}^-$ &
2 &   10.74 $\pm$ 0.04 & 10.63 &  10.6 $\pm$ 0.1 &
10.69 $\pm$ 0.03  \\
 & $\frac{3}{2}^+$ &
1 &   8.39 $\pm$ 0.15 & 8.84 &  8.78 $\pm$ 0.08 &
8.56 $\pm$ 0.07 \\
 & $\frac{9}{2}^+$ &
5 &   222 $\pm$ 3 & 228 &  229 $\pm$ 2 &
223 $\pm$ 2  \\
\hline
%\multicolumn{7}{l}{$^{2c}$ - two-cluster model} \\
%\multicolumn{7}{l}{$^{4c}$ - four-cluster model }\\
\end{tabular}
\end{center}
\label{table3}
\end{table*}

%%%%%%%%%%%%%%%%%%%%%%%%%%%%%%%%%%%%%%%%%%%%%%%%%%%%%%%%%%%%%%%%%%%%%%%%%

\subsection{Mirror ANCs in a microscopic cluster model}

%%%%%%%%%%%%%%%%%%%%%%%%%%%%%%%%%%%%%%%%%%%%%%%%%%%%%%%%%%%%%%%%%%%%%%%%%

The relation (\ref{rna}) for mirror ANCs obtained in the two-body model 
can be extended 
to many-body systems. The expression for an ANC in 
the many-body case is \cite{Tim98}
\beq
C^{(i)}_l = - \frac{2\mu}{\hbar^2} \int_0^{\infty} dr \, r^2
\tilde{\phi}_l^{(i)}(r) \, \la [\Phi_{X_i}^{J_{X_i}} \otimes 
Y_{l}(\hat{r})]_{J_A}
\eol
\times \Phi_{\alpha} || V_N + V_{C_i} - \tilde{V}_i || \Psi_A^{J_A}\ra
\eeqn{mbanc}
where $\Psi_A^{J_A}$, $\Phi_{\alpha}$ and $\Phi_{X_i}^{J_{X_i}}$ are 
the many-body wave 
functions of the nucleus $A$, $\alpha$-particle and the decay product
$X_i$, and $J_{A}$ and $J_{X_i}$ are the total spins of $A$ and $X_i$.
The integration in the source term 
$\la [\Phi_{X_i}^{J_{X_i}} \otimes Y_{l}(\hat{r})]_{J_A} \Phi_{\alpha}
|| V_N + V_{C_i} - \tilde{V}_i || \Psi_A^{J_A}\ra$
is carried out over the internal   coordinates of $\alpha$ and $X_i$
and the potentials $V_N$ and $V_C$ are the sums of the two-body nuclear
and Coulomb interactions. Following the reasoning of   section A,
we get the formula (\ref{rna}). The deviation from this formula
will be determined by the remainder terms $R_C(a)$, $R_{\Delta\Psi}$,
$R_B(a)$,  $R_{\delta\phi}(a)$ and $R_{\Delta V_C}(a)$ defined by
Eqs. similar to (\ref{RC}),   (\ref{RDeltaPsi}),  (\ref{Rdeltaphi}),
(\ref{RB}) and (\ref{RDeltaVC}) but in the
integrands of which   $V\Psi$  is be replaced by
the matrix elements of the 
$\la [\Phi_{X_i}^{J_{X_i}} \otimes Y_{l}(\hat{r})]_{J_A}\Phi_{\alpha}
|| V|| \Psi_A^{J_A}\ra$ type.

The main difference between the two-body and many-body cases is that 
$V_C - V_{C_0}$  is not zero   at $r > R_N$.
It contains long range contributions from the $r^{-\lambda}$ 
($\lambda \geq 2$) terms
the strengths of which are determined by the matrix elements
$\la [\Phi_{X_i}^{J_{X_i}} \otimes Y_{l}(\hat{r})]_{J_A}\Phi_{\alpha}
|| M(E\lambda) || \Psi_A^{J_A}\ra$ where $M(E\lambda)$ is the 
electromagnetic operator   of multipolarity $\lambda$ \cite{Tim05a}.
If these matrix elements are large, then all the remnant terms  
that contain $\Delta V_{C_i}$
may cause significant differences between ${\cal R}$ and ${\cal R}_0$.
This is expected for nuclei with strongly deformed and/or easily excited
cores.

Another factor that may lead to additional differences
between ${\cal R}$ and ${\cal R}_0$ in many-nucleon systems is that
the condition $\Psi_l^{(1)} \approx \Psi_l^{(2)}$ for the validity of
Eq. (\ref{rna}) in the two-body case is
replaced by the equality of the projections 
$\la [\Phi_{X_i}^{J_{X_i}} \otimes Y_{l}(\hat{r})]_{J_A}\Phi_{\alpha}
| \Psi_A^{J_A}\ra$ (or overlap integrals)
of the mirror wave functions for nuclei
$^A_NZ$ and $^A_ZN$
into the mirror channels $X_i + \alpha$. 
If the norms of these overlap integrals (or spectroscopic factors) differ
then the terms $R_{\Delta\Psi}$, $R_{\delta\phi}(a)$ 
and $R_{\Delta V_C}(a)$ will increase. This can be especially
important for weak components of overlap integrals where symmetry
breaking in the spectroscopic factors may become large.

Our previous study of many-body effects in mirror virtual nucleon decays 
suggests that they are on average of the order of 7$\%$ \cite{Tim05a},
although stronger deviations in some
individual cases were observed as well.
Here, we study the many-body effects in mirror $\alpha$-particle ANCs
using a multi-cluster model of the same type as in Ref. \cite{Tim05a}
for the same mirror pairs $^7$Li-$^7$Be, $^{11}$B-$^{11}$C and 
$^{19}$F-$^{19}$Ne considered above in the two-body model.

The multi-channel cluster wave function for a nucleus $A$ consisting of 
a core $X$ and an $\alpha$-particle can be represented as follows:
\beq
\Psi_A^{J_AM_A} = \sum_{l \omega  J_{X}} {\cal A} \, 
\Phi_{\alpha}\left[
g^{J_{X}J_A}_{\omega l}(\ve{r}) \otimes \Phi_{X}^{J_{X}}
\right]_{J_AM_A}
\eeqn{wfMCM}
where ${\cal A}$ is the antisymmetrization operator which permutes
nucleons between the $\alpha$-particle and the core.
Both the $\alpha$-particle wave function and 
the ``core" wave  function $\Phi_{X}^{J_{X}}$  corresponding to the  total
spin $J_{X}$ are defined in the
translation-invariant harmonic-oscillator shell model.  
In addition, for $^{11}$C we used the three-cluster model of Ref. 
\cite{Des95}
in which
$\Phi_X^{J_X}$ is defined in a two-cluster model.
The quantum number $l$ labels the orbital momentum of 
the $\alpha$-particle.
The relative wave function $g^{J_XJ_A}_{\omega lm}(\ve{r}) =
g^{J_XJ_A}_{\omega l}(r)Y_{lm}(\hat{\ve{r}})$
is determined using the microscopic R-matrix method 
\cite{Des90} to provide
the correct asymptotic behaviour
\beq
g^{J_XJ_A}_{\omega l}(r) \approx C_{ l,\omega }^{J_XJ_A}
\frac{W_{-\eta,l+1/2}(2\kappa r)}{r},
 \,\,\,\,\,\, r \rightarrow \infty,
\eeqn{asg}
%where $\kappa = \sqrt{2\mu S_{\alpha}}/\hbar$, $\mu$ is the reduced mass
%for the $X+\alpha$ system, $S_{\alpha}$ is the $\alpha$-particle separation
%energy, $\eta = Z_XZ_{\alpha}e^2\mu/\hbar^2\kappa$, 
determined by the  Whittaker function and the ANC
$C_{ l,\omega }^{J_XJ_A}$.

\begin{table*} 
\caption{
The range of changes in squared ANCs
$C^2_{\alpha}(2)$ and $C^2_{\alpha}(1)$ (in fm$^{-1}$) for mirror nuclei 2 and 1
($Z_2 > Z_1$) and in their ratio ${\cal R}_{MCM}$ with the choice of oscillator
radius and the NN potential. For $^{11}$B - $^{11}$C, this range include also
changes with different number of clusters. Also shown are the
spectroscopic factors $S_{\alpha}(2)$, $S_{\alpha}(1)$ and their ratio
${\cal R}_S^{MCM}=S_{\alpha}(2)/S_{\alpha}(1)$.
} 
\begin {center}
\begin{tabular}{   p{1.0 cm}
p{1.0 cm} p{2.8 cm} p{2.8 cm} p{3.2 cm}  p{2.2 cm} p{2.2 cm} 
p{1.6 cm} }
\hline 
\hline
 $J^{\pi}$ & $l$ & $C^2_{\alpha}(2)$ & $C^2_{\alpha}(1)$ & 
 ${\cal R}_{MCM}$ & $S_{\alpha}(2)$ & $S_{\alpha}(1)$ & ${\cal R}_S^{MCM}$ \\
 
\hline
\\

\multicolumn{8}{c}{$^{7}{\rm Li} - ^{7}{\rm Be}$} \\ \\

$3/2^-$ & 1 & 19.4-30.4 & 14.3-22.6 & $1.35\pm 0.01$ & 
1.13-1.15 & 1.14-1.16 & 0.99-1.00\\
$1/2^-$ & 1 & 14.9-22.7 & 10.4-16.0 & $1.43\pm 0.01$ & 
1.12-1.14 & 1.13-1.15 & 0.99\\

\\
\multicolumn{8}{c}{$^{11}{\rm B} - ^{11}{\rm C}$} \\ \\
$3/2^-_1$ & 0 & (0.54-2.15)$\times 10^4$ & (0.34-1.25)$\times 10^4$ &
$1.65\pm 0.07$ & 0.29-0.38 & 0.28-0.37 & 1.02-1.03 \\
  & 2 & (1.69-6.74)$\times 10^3$ & (1.12-4.26)$\times 10^3$ &
$1.54\pm 0.05$ & 0.45-0.51 & 0.44-0.51 & 1.00-1.01 \\
$1/2^-$ & 2 & (0.69-3.69)$\times 10^3$ & (0.42-2.19)$\times 10^3$ 
& $1.66\pm 0.03$ & 0.37-0.42 & 0.37-0.41 & 1.00-1.01\\
$5/2^-$ & 2 & (0.51-2.19)$\times 10^3$ &  (0.28-1.12)$\times 10^3$ &
$1.90\pm 0.07$ & 0.64-0.76 & 0.64-0.75 & 1.01-1.02\\
$3/2^-_2$ & 0 & (0.76-1.40)$\times 10^3$ & 338-612 & $2.25\pm 0.04$ & 0.09-0.15 &
0.09-0.15 & 0.98-1.00\\
 & 2 & 41.9-428 & 19.4-191 & $2.18\pm 0.06$ & 0.1-0.23 & 0.09-0.22 & 1.04-1.06\\
$1/2^+$ & 1 & (0.47-3.00)$\times 10^4$ &  (1.04-6.41)$\times 10^3$ & 
$4.59\pm 0.05$ & 0.66-0.88 & 0.66-0.87 & 1.00-1.01\\
$7/2^-$ & 2 & 4.67-20.8 & 1.08-4.67 & $4.39\pm 0.07$ & 0.014-0.026 & 
0.013-0.025 & 1.03-1.05\\
 & 4 & 0.20-0.75 & 0.08-0.30 & $2.59\pm 0.10$ & 0.07-0.34 & 0.06-0.33 & 1.01-1.02\\
$5/2^+$ & 1 &  (1.0-5.0)$\times 10^4$ & (0.75-3.68)$\times 10^3$ &
$13.4\pm 0.2$ & 0.84-0.95 & 0.83-0.94 & 1.00-1.01\\
 & 3 & 13.3-243 & 1.72-28.9 & $8.03\pm 0.36$ & 0.034-0.064 & 0.033-0.059 
 & 1.01-1.08\\
$3/2^+$ & 1 &(0.30-1.2)$\times 10^{15}$ & 179-703 & $(1.68\pm 0.02)\times 10^{12}$&
0.16-0.38 & 0.16-0.38 & 1.00-1.01\\
& 3 & (0.09-1.67)$\times 10^{13}$ & 2.47-44.1 & $(3.60\pm 0.07)\times 10^{11}$ &
0.094-0.18 & 0.093-0.18 & 1.00-1.03\\
\\

\multicolumn{8}{c}{$^{19}{\rm F} - ^{19}{\rm Ne}$} \\ \\
$1/2^+$ & 1 & (0.42-1.24)$\times 10^7$ & (1.04-3.00)$\times 10^6$ &
$4.12\pm 0.06$ & 0.17-0.55 & 0.17-0.56 & 0.98-1.00\\
$5/2^+$ & 3 & (2.39-7.97)$\times 10^5$ & (0.57-1.89)$\times 10^5$ & 
$4.23\pm 0.07$ & 0.15-0.41 & 0.16-0.42 & 0.98-1.00\\
$1/2^-$ & 0 & (1.38-5.07)$\times 10^7$ &
(0.30-1.08)$\times 10^7$ & $4.69\pm 0.02$ & 0.42-0.69 & 0.42-0.69 & 1.00-1.01\\
$5/2^-$& 2 & (0.71-2.81)$\times 10^7$ &
(0.75-2.92)$\times 10^6$  & $9.58\pm 0.04$ & 0.42-0.68 & 0.41-0.67 & 1.00-1.01\\
$3/2^+$ & 1 & (1.34-4.59)$\times 10^7$ & 
(1.63-5.38)$\times 10^6$ & $8.39\pm 0.15$ & 0.17-0.67 & 0.18-0.68 & 0.97-0.99 \\
$3/2^-$ & 2 & (0.88-3.43)$\times 10^7$ &
(0.82-3.18)$\times 10^6$ & $10.74 \pm 0.04$ & 0.44-0.69 & 0.44-0.69 & 1.00-1.01\\
$9/2^+$ & 5 & (1.31-6.39)$\times 10^5$  &
(0.60-2.87$\times 10^3$ & $222 \pm 3$ & 0.16-0.30 & 0.16-0.30 & 0.99\\
\\

  \hline
  \hline
 \end{tabular}
\end{center}
\nonumber
\label{table1}
\end{table*}

The MCM requires some choice of the oscillator radius $b$ to describe
the internal structure of the clusters. In all three mirror pairs considered
in this paper, the oscillator radius that  provides a good description
of the $\alpha$-particle  differs
significantly from that of the core. Dealing with different
$b$ for each of the cluster would create big difficulties in
using the MCM. Therefore, we use the same value of $b$ 
for both clusters but do the calculations twice.
The first time we use $b$ =  1.36 fm that reproduces the r.m.s. 
radius of the $\alpha$-particle and minimises its binding energy, and
the second time we use either $b$ = 1.5 fm 
(to describe the triton and/or $^3$He core
for the $^7$Li - $^7$Be mirror pair) or 
$b$ = 1.6 fm (for $^{11}$B - $^{11}$C
and $^{19}$F - $^{19}$Ne).
Our previous calculations for $^{17}$O - $^{17}$F have shown that different 
oscillator radii change strongly the absolute value of neutron and proton ANCs
but does not change  their ratio very much \cite{Tim05a}.
In the three-cluster calculations for the
$^{11}$B - $^{11}$C mirror pair we used only one value of the oscillator 
radius, $b$ = 1.36 fm,  the same as in Ref. \cite{Des95}.

\begin{figure*}[t]
\centerline{\psfig{figure=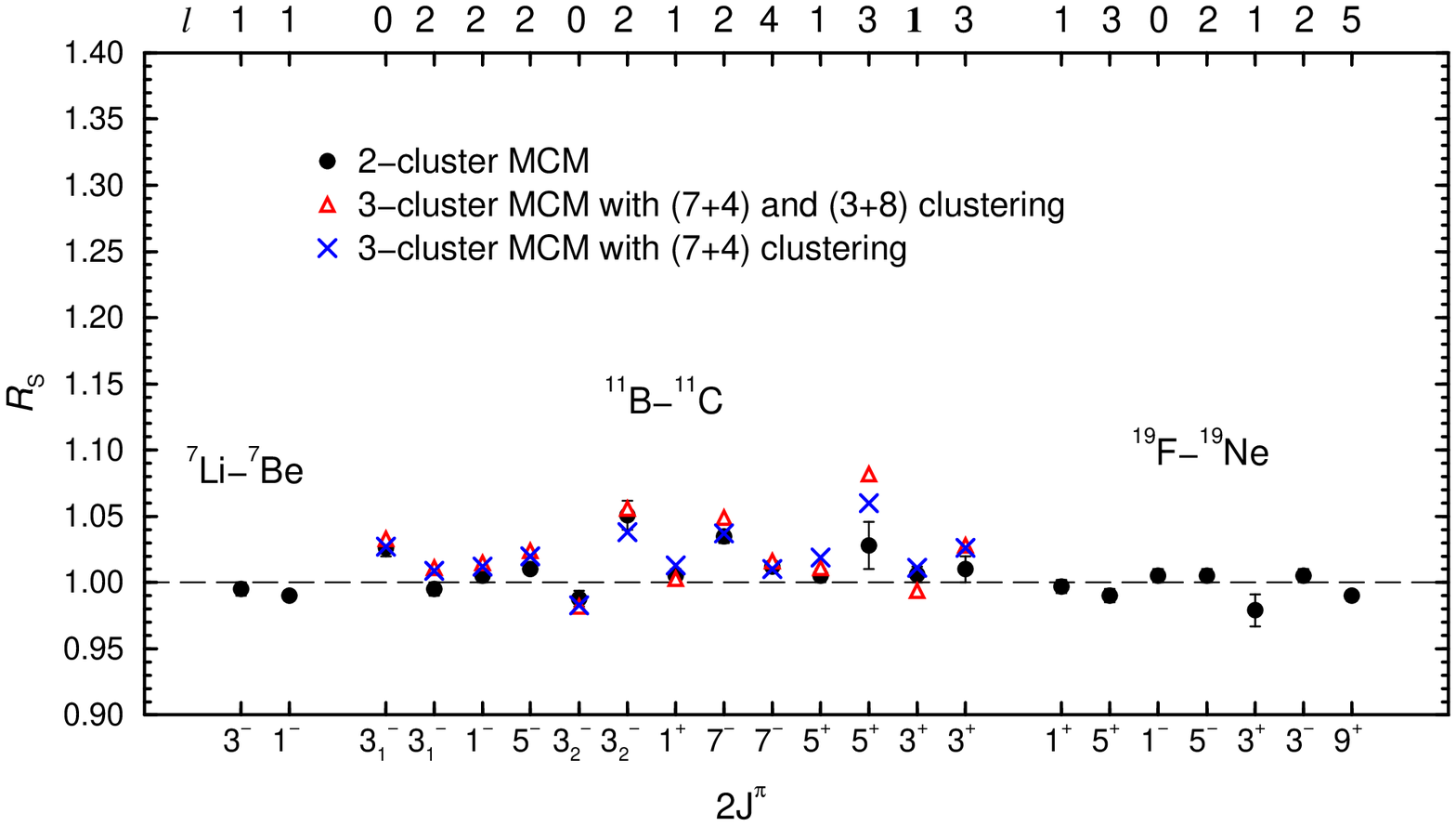,width=0.86\textwidth} }
\caption{ Ratio ${\cal R}_S$
of the $\alpha$-particle spectroscopic factors 
in the $^{7}$Li-$^{7}$Be, $^{11}$B-$^{11}$C and $^{19}$F-$^{19}$Ne
mirror pairs, calculated in the two-cluster (filled circles)
and three-cluster microscopic cluster models in which both
the $^7$Li+$\alpha$  ($^7$Be+$\alpha$)
and t+$^8$Be ($^3$He+$^8$Be)  (triangles) or only
the $^7$Li+$\alpha$  ($^7$Be+$\alpha$)
 partitions (crosses) have been taken into account.
}
\end{figure*}

For each oscillator radius, we use two NN potentials,
the  Volkov potential V2 \cite{volkov}
and the Minnesota (MN) potential \cite{minnesota}, except in
three-cluster calculations for $^{11}$B-$^{11}$C where only V2 is used.
The two-body spin-orbit force \cite{BP81}
 with $S_0$ = 30 MeV$\cdot$fm$^5$
 and the Coulomb interaction are also included.
Both   V2 and MN have one adjustable parameter that gives the
strength of the odd NN potentials $V_{11}$ and $V_{33}$. We fit this
parameter  in each  case
to reproduce the experimental values for the $\alpha$-particle
 separation energies.
 Slightly different adjustable parameters in mirror nuclei, needed to
reproduce these energies,
simulate charge symmetry breaking  of the effective NN interactions,
which could be a consequence of  charge symmetry breaking in 
realistic NN interactions. 

The range of changes in squared ANCs $C_{\alpha}^2(2)$ and $C_{\alpha}^2(1)$
in mirror nuclei 2 and 1 is given in Table II. Similar to previous studies of
of one-nucleon ANCs in Refs. \cite{BT92,Tim05a,Tim06a}, the V2 potential
gives larger $C_{\alpha}^2$  values than the MN (up to a factor of two) at 
a fixed oscillator radius $b$ and the different choices of $b$ give 
a comparable
change (up to the factor of two) in $C_{\alpha}^2$ at a fixed NN potential.
The range of change in the  ratio
${\cal R}_{MCM}$ with different choices of oscillator radius and the 
NN potential
are also given in Table II. For $^{11}$B - $^{11}$C, this range  includes 
changes with different number of clusters.  In Table I the average value of
${\cal R}_{MCM}$ is compared to the analytical estimate ${\cal R}_0$ and
to predictions within the potential model ${\cal R}_{PM}$. 
To visualise the deviation between
${\cal R}_{MCM}$ and ${\cal R}_0$  we plot the ratio 
${\cal R}_{MCM}/{\cal R}_0$ in Fig.2. 
%In all cases ${\cal R}_0$
%has been calculated assuming that $R_N = a_m$.

We have also calculated the $\alpha$-particle spectroscopic factors 
$S_{\alpha}$ defined as
\beq
S_{\alpha} = \left(\begin{array}{c} A \\4\end{array}\right) 
\int_0^{\infty} dr \, r^2
\left|\la [\Phi_{X_i}^{J_{X_i}} \otimes Y_{l}(\hat{r}) ]_{J_A}\Phi_{\alpha}
| \Psi_A^{J_A}\ra\right|^2
\eol
\eeqn{SF}
and have shown their range of variation in Table II. The ratio 
${\cal R}_S^{MCM} = S_{\alpha}(2)/S_{\alpha}(1)$
of these spectroscopic factors is given in Table II as well and is plotted
in Fig.3. We also 
calculate the ratio ${\cal R}_{b_{\alpha}}^{MCM} = (b_{\alpha}(2)/
b_{\alpha}(1))^2$ of the normalized squared ANCs 
$b_{\alpha} = C_{\alpha}/\sqrt{S_{\alpha}}$. As in  the case of
mirror virtual nucleon decays studied in Ref.
\cite{Tim05a, Tim06b}, 
the approximate equality  ${\cal R}_{b_{\alpha}}^{MCM}\approx {\cal R}_{PM}$
means that in mirror nuclei the effective local
nuclear $\alpha$-core interaction 
can be considered to be the same.

We now discuss individual mirror pairs in more details.

{\bf $^7$Li - $^7$Be}. The squared ANCs in these mirror nuclei change 
by about 55$\%$ with different oscillator radii and NN potentials. However, 
the ratio $C_{\alpha}(^7$Be$)/C_{\alpha}(^7$Li) changes by only
about 1.5$\%$ both in the 
ground and the first excited states. This ratio differs from the analytical
estimate ${\cal R}_0$ 
by no more than 3$\%$ and 4$\%$ for the ground and the first excited
state respectively and agrees reasonably well with the potential model
calculations.  
The mirror symmetry in spectroscopic factors is also clearly seen.
Some minor differences in 
${\cal R}_{b_{\alpha}}^{MCM}$ and ${\cal R}_{PM}$ are present
which means that the effective local nuclear $t+\alpha$ and $^3$He + $\alpha$
interactions differ slightly.
Since the $^7$Li and  $^7$Be ANCs determine the cross sections
for the $^3$H($\alpha,\gamma$)$^7$Li and $^3$He($\alpha,\gamma$)$^7$Be 
capture
reactions at zero energies, the mirror symmetry of the $\alpha$-particle ANCs
means that relations should exist between the astrophysical $S$-factors
of these reactions. Thus, with our value of ${\cal R}_{MCM}$ the ratio
$S_{34}(^7$Be$)/S_{34}(^7$Li) at zero energy is 6.6 and 5.9 for the ground 
and the first excited states respectively.

{\bf $^{11}$B - $^{11}$C}. The calculations for this mirror pair have been 
performed for all excited states that are below the $\alpha$-particle emission
threshold in $^{11}$C. In the two-cluster model, 
only the ground and the $\frac{1}{2}^-$ first excited state 
in the $^7$Li - $^7$Be mirror cores have been taken into account. 
In the three-cluster model, both the $^7$Li+$\alpha$  ($^7$Be+$\alpha$)
and t+$^8$Be ($^3$He+$^8$Be) partitions are taken into account with
the first excited states $\frac{1}{2}^-$,$\frac{3}{2}^-$,$\frac{5}{2}^-$
and $\frac{7}{2}^-$ in $^7$Li ($^7$Be) and the first 0$^+$ and 2$^+$ states
in $^8$Be included \cite{Des95}.

The squared ANCs  calculated in the two-cluster MCM
change with different NN potential and oscillator radius choice  
by the factor of  four on average (see Table II). Taking two-cluster nature
of $^7$Li and $^7$Be into account in most cases significantly increases ANCs 
thus increasing the range of their variations with model assumptions.
However, in all cases
the ratio ${\cal R}_{MCM}$ changes by no more than 9$\%$.
The ${\cal R}_{MCM}$ values
obtained in the two-cluster model %are smaller than that obtained in the 
%three-cluster model and on average they 
are close to the analytical estimate 
${\cal R}_0$ and to
the potential model prediction ${\cal R}_{PM}$, agreement being
within 1-5$\%$ (see Fig.2). For the second
$\frac{3}{2}^-$ state with  $l=2$, a larger deviation from
${\cal R}_0$ and ${\cal R}_{PM}$   (5-10$\%$) 
coincides with larger symmetry breaking in the
mirror spectroscopic factors (see Fig.3). 

The ${\cal R}_{MCM}$ values obtained in the three-cluster MCM are 
significantly larger than the predictions of the two-cluster model. 
This is caused mainly by the influence of the 
t+$^8$Be and $^3$He+$^8$Be  channels. When these channels are removed,
so that only the $^7$Li+$\alpha$  and $^7$Be+$\alpha$ partitions
are left, then both the two-cluster and three-cluster MCM predict
very similar results for the ratio  ${\cal R}_{MCM}$ (see Fig.2). At the
same time, the ratio of mirror spectroscopic factors is not very much 
influenced by the t+$^8$Be ($^3$He+$^8$Be) clustering, although  for the 
$\frac{5}{2}^+$ state with $l$ = 3 it is  slightly reduced.
This happens because the effective local $\alpha - ^7$Li and
$\alpha - ^7$Be interaction differ. This can be seen by comparing the
${\cal R}_{b_{\alpha}}^{MCM}$ obtained in the three-cluster calculations
with ${\cal R}_{PM}$. In two-body calculations these quantities
agree with each other within the uncertainties of their calculations
for most of the mirror states.

{\bf $^{19}$F - $^{19}$Ne}. The two-cluster
MCM calculations for this mirror pair have been 
performed for all excited states that are below the $\alpha$-particle emission
threshold in $^{19}$Ne. The mirror cores $^{15}$N - $^{15}$O were
considered both in the ground and the first excited state $\frac{3}{2}^-$.  
We have found that different choices of the oscillator radius strongly
influence  the mixture of the $\alpha$+$^{15}$N($\frac{1}{2}^-$)
and $\alpha$+$^{15}$N($\frac{3}{2}^-$) configurations in all the 
states of $^{19}$F, leading to large 
changes in spectroscopic factors and ANCs.
The same is true for the $\alpha$+$^{15}$O($\frac{1}{2}^-$)
and $\alpha$+$^{15}$O($\frac{3}{2}^-$) configurations in $^{19}$Ne. However,  
despite the 3-5 times change in squared ANCs, the ratio 
${\cal R}_{MCM}$ of mirror squared
ANCs changes by less 3.5$\%$. This ratio is close to both the analytical 
estimate ${\cal R}_0$ and   the predictions of the potential model
${\cal R}_{PM}$. The deviation between ${\cal R}_{MCM}$ and these estimates
does not exceed 5$\%$. The mirror symmetry in spectroscopic factors is
also clearly seen. In most cases ${\cal R}_b^{MCM}$ and ${\cal R}_{PM}$
agree within uncertainties of their definition which means that
mirror symmetry in the effective local $\alpha + ^{15}$N and 
$\alpha + ^{15}$O interactions is a good assumption.\\

%%%%%%%%%%%%%%%%%%%%%%%%%%%%%%%%%%%%%%%%%%%%%%%%%%%%%%%%%%%%%%%%%%%%%%%%%%

\section{Bound-unbound mirror pairs}

%%%%%%%%%%%%%%%%%%%%%%%%%%%%%%%%%%%%%%%%%%%%%%%%%%%%%%%%%%%%%%%%%%%%%%%%%% 

\begin{table*}[t]
\caption{ 
Range of change for the width $\Gamma_{\alpha}$ (in MeV)
of an $\alpha$-particle resonance
and for its mirror squared ANC $C^2_{\alpha}$ (in fm$^{-1}$)
with different model parameters.
The results of calculation are given both in
the potential model and in the MCM.
 } 
\begin {center}
\begin{tabular}{ p{2.4 cm} p{1 cm} p{1 cm} p{3 cm}
p{3 cm} p{3 cm} p{3  cm} }
\hline 
\\ 
 & & & \multicolumn{2}{c}{Potential model} & 
  \multicolumn{2}{c}{Microscopic cluster model} \\
Mirror pair  &  $J^{\pi}$ & 
$l$  &
 $\Gamma_{\alpha}$  & $C^2_{\alpha}$ & $\Gamma_{\alpha}$ & $C^2_{\alpha}$ \\
 \hline
$^{11}{\rm B}  - ^{11}{\rm C}$ & $\frac{3}{2}^-_3$ & 
0 &   (2.13 - 3.53) $\times 10^{-3}$ & (2.04 - 3.40 ) $\times 10^{6}$ &
(0.98 - 2.51) $\times 10^{-3}$ & (8.91 - 25.3 ) $\times 10^{5}$ \\
& & 2 &  (1.20 - 2.44 ) $\times 10^{-4}$ & (8.17 - 16.3 ) $\times 10^{4}$
& (3.25 - 11.2 ) $\times 10^{-5}$ & (2.18 - 8.11 ) $\times 10^{4}$\\
$^{19}{\rm F}  - ^{19}{\rm Ne}$ & $\frac{3}{2}^+_2$ & 
1 &   (3.95 - 10.2 ) $\times 10^{-10}$ & (1.23 - 3.11 ) $\times 10^{23}$
& (0.76 - 2.58 ) $\times 10^{-10}$ & (2.21 - 7.54 ) $\times 10^{22}$\\
 & $\frac{7}{2}^-_2$ &  
4 &   (3.67 - 15.1 ) $\times 10^{-10}$ & (4.68 - 18.4 ) $\times 10^{72}$
& (0.98 - 3.40 ) $\times 10^{-13}$ & (2.54 - 23.8 ) $\times 10^{69}$\\
\hline
\end{tabular}
\end{center}
\label{table3}
\end{table*}

%Since isospin symmetry is present  in mirror virtual $\alpha$-decays
%and  because     it exists in mirror
%virtual neutron and real proton decays
%it is reasonable to expect this symmetry also
%in mirror states one of which is below and another is above 
%the $\alpha$-emission threshold. 
The symmetry in mirror $\alpha$-decays can   be extended to   
bound-unbound mirror pairs.
As in the case of nucleon decays \cite{Tim03,Tim05b},
such a symmetry would manifest itself as a link between the ANC of the
bound $\alpha$-particle state
and the width   of its analog resonant state. This follows from the
possibility to represent the resonance width by an integral
similar to (\ref{anc}) and (\ref{mbanc}).
For isolated narrow resonances, the generalization of Eq. (17)
of Ref. \cite{Tim05b} for the two-body  $\alpha$-particle case 
gives  the width $\Gamma_{l}^0$ as
\beq
\Gamma_{l}^0 \approx \frac{2\kappa_R}{E_R} 
 \left| \int_0^{R_{m}}
dr \, r F_l(\kappa_R r) (V_N - \Delta V_C )
 \Psi_{l}^{BSA}(r) \right|^2,
 \eol
\eeqn{gamma}
where $E_R$ is the resonance energy, $k_R = \sqrt{2\mu E_R/\hbar^2}$,
$F_l$ is the regular Coulomb wave function and $\Psi_{l}^{BSA}$ is
a wave function of the $\alpha$-particle resonance 
in  the bound-state approximation. This function
%$\Psi_{l}^{BSA}$
has the dimension  of a bound-state wave function and is defined and
normalized within 
some channel radius $R_{m}$ taken well outside the range of 
the $\alpha$-core interaction. The  width $\Gamma_{l}^0$
defined by Eq. (\ref{gamma}) is related to the residue $\gamma_l^2$
at the R-matrix pole by \cite{Lan58},
\beq
\Gamma_{l}^0 = 2 \kappa_R R_{m}\,  \gamma_l^2 /| O_l(\kappa_R R_{m})|^2,
\eeqn{residue}
where $O_l$ is the outgoing Coulomb function. It determines
the observable width $\Gamma_{l}$  by 
\beq
\Gamma_{l}  =\Gamma_{l}^0 /(1+\gamma_l^2S'_l)^{-1},
\eeqn{obsgamma}
where $S_l = {\rm Re} (\kappa_R R_{m}O'_l/O_l)$ and the derivation
is performed with over the energy $E$. For very narrow
resonances, such that $\gamma_l^2S'_l \ll 1$, the observed
width, $\Gamma_{l}$, and the one related to the residue in the
R-matrix pole, $\Gamma_{l}^0$, are the same. It is for such cases
that the analytical expression for the ratio
\beq
{\cal R}_{\Gamma} = \Gamma_{\alpha}/C^2_{\alpha}
\eeqn{}
can be derived. Following the reasoning of Sec. II.A  we get
the approximate model-independent formula
\beq
{\cal R}_{\Gamma} \approx   {\cal R}_0^{res} = \frac{\hbar^2 k_R}{\mu}
\frac{\varepsilon_{b.s.}}{E_R} \left| \frac{F_l(k_R R_N)}
{F_l(i\kappa_{b.s.} R_N)}\right|^2
\eeqn{rnres}
where $ \varepsilon_{b.s.}$ is the binding energy of a bound
$\alpha$-particle state and
$\kappa_{b.s.}= \sqrt{2\mu \varepsilon_{b.s.}}/\hbar$. 
As in the case of bound mirror pairs,
the difference  between ${\cal R}_0^{res}$ and the exact value of
${\cal R}_{\Gamma}$
will be determined by   remainder terms similar
to those given in Eqs. (\ref{RC}), (\ref{RDeltaPsi}), (\ref{Rdeltaphi}),
(\ref{RB}) and (\ref{RDeltaVC}),
and their magnitude will depend on how
similar are the bound state $\alpha$-particle wave
function and its mirror analog
$\Psi_{l}^{BSA}$. As for bound mirror pairs,  
the formula (\ref{rnres})
will be more accurate if the function
$ | F_l(k_R r)/F_l(\kappa_{b.s.}r)| $ varies slowly near
$r \approx R_N$. This function changes the most slowly
near its maximum, at $r = a_m$.

\begin{figure}[t]
%\centerline{\psfig{figure=../../alpha.BU.eps,width=0.40\textwidth} }
\centerline{\psfig{figure=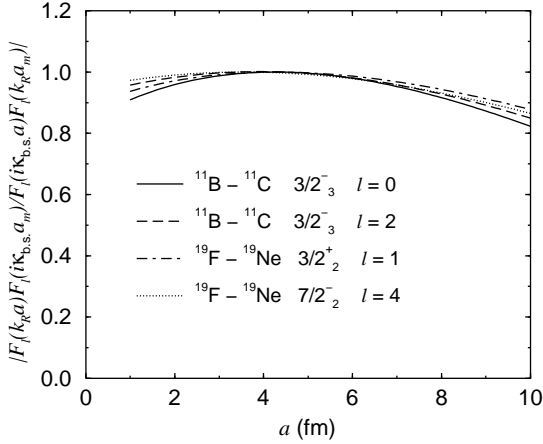,width=0.40\textwidth} }
\caption{ The ratios
$ | F_l(k_R a)F_l(\kappa_{b.s.}a_m)/F_l(\kappa_{b.s.}a) F_l(k_R a_m)| $  
as a function of $a$.
}
\end{figure}

\begin{table*}[t]
\caption{ 
Analytical estimate  ${\cal R}_{PM}^{res}$  and the two-body
(${\cal R}_{0}^{res}$) and MCM prediction (${\cal R}_{\Gamma}^{MCM}$)
for the ratio ${\cal R}_{\Gamma}$ (all in MeV$\cdot$fm) for
some mirror states in $^{11}$B-$^{11}$C and $^{19}$F-$^{19}$Ne.
  $\varepsilon_{b.s}$ is the binding energy of a bound $\alpha$-particle
  state and $E_R$ is the resonance energy of its mirror analog while
  $l$ is the orbital momentum.
 } 
\begin {center}
\begin{tabular}{ p{2.4 cm} p{1 cm} p{1.4 cm} p{1.4 cm}
p{1 cm} p{2.5 cm} p{3.5 cm} p{3 cm}}
\hline 
\\ 
Mirror pair  &  $J^{\pi}$ &  $\varepsilon_{b.s}$ & $E_R$ &
$l$  
&
 ${\cal R}_0^{res}$  &  ${\cal R}_{PM}^{res}$
 &  
 ${\cal R}_{\Gamma}^{MCM}$  \\
 \hline
$^{11}{\rm B}  - ^{11}{\rm C}$ & $\frac{3}{2}^-_3$ & 0.104 & 0.5615 &
0 &   $1.18 \times 10 ^{-9}$ & $1.05  \times 10 ^{-9}$ &  
(1.05 $\pm$ 0.06) $\times 10 ^{-9}$ \\
&  &  &   &
2 &   $1.52 \times 10 ^{-9}$ & $(1.48 \pm 0.01)  \times 10 ^{-9}$ &  
(1.47 $\pm$ 0.03) $\times 10 ^{-9}$ \\
$^{19}{\rm F}  - ^{19}{\rm Ne}$ & $\frac{3}{2}^+_2$ & 0.1056 & 0.5038 &
1 &   $3.30 \times 10 ^{-33}$ & $(3.25 \pm 0.04)  \times 10 ^{-33}$ &  
(3.42 $\pm$ 0.04) $\times 10 ^{-33}$ \\
 & $\frac{7}{2}^-_2$ & 0.0151 & 0.6677 &
4 &   $7.86 \times 10 ^{-84}$ & $(8.00 \pm 0.18)   \times 10 ^{-84}$ &  
(1.32 $\pm$ 0.12) $\times 10 ^{-83}$ \\
\hline
\end{tabular}
\end{center}
\label{table4}
\end{table*}

In Fig. 4 we plot the function 
$ | F_l(k_R a)F_l(\kappa_{b.s.}a_m)/$ $F_l(\kappa_{b.s.}a) F_l(k_R a_m)| $ 
for  three mirror pairs of excited states
$^{11}$B($\frac{3}{2}^-_3$, 8.560 MeV) - 
$^{11}$C(($\frac{3}{2}^-_3$, 8.105 MeV),
$^{19}$F($\frac{3}{2}^+_2$, 3.908 MeV) - 
$^{19}$Ne($\frac{3}{2}^+_2$, 4.033 MeV) and
$^{19}$F($\frac{7}{2}^-_2$, 3.999 MeV) -
$^{19}$F($\frac{7}{2}^-_2$, 4.197 MeV).
The $\alpha$-particle in the chosen states of $^{11}$B and $^{19}$F 
is weakly bound and its mirror states  in $^{11}$C and $^{19}$Ne are
resonances which are important for some astrophysical applications. This
ratio is almost a constant for $r \sim 4 - 6$ fm which is close to $R_N$.

We compare ${\cal R}_0^{res}$, calculated assuming  $R_N = a_m$,
to  ${\cal R}_{\Gamma}$ obtained in exact two-body calculations.
To perform the two-body calculations,
 we have chosen an $\alpha$-core potential
of the Woods-Saxon form and varied its diffuseness from 0.35 fm to 0.95 fm.
For each diffuseness the depth and the radius of this
potential  were adjusted
to reproduce simultaneously both the $\alpha$-particle
separation energy $\varepsilon_{b.s.}$ in a chosen state
and the position $E_R$ of the resonance in its mirror analog. 
The width has been determined from the behaviour of the
resonant phase shift $\tan \delta_{l} =  \Gamma_{l}(E)/2(E-E_R)$ near
$E_R$. The range of change in squared ANCs and in resonance widths
with the potential geometry is presented in Table III. The  widths
change by a factor from 1.65 to 4.1 and
the ANCs squared in the mirror states change by the same amount so that
${\cal R}_{PM}^{res}$ changes by less than 2$\%$ with respect to an average
value.   These average values are very close to  
${\cal R}_{0}^{res}$ when $l_{\alpha} \ne 0$ (see Table IV).
In the $l_{\alpha} = 0$ case, when the centrifugal barrier in absent,
the approximation (\ref{rnres}) becomes less accurate, with
 ${\cal R}_{PM}^{res}$ being smaller than
${\cal R}_{0}^{res}$ by 12$\%$. This loss of accuracy 
is probably caused by a larger difference in mirror $s$-wave functions
when one of the $\alpha$-particles is loosely-bound. 
In all cases, 
the agreement between ${\cal R}_{PM}^{res}$
and ${\cal R}_{0}^{res}$ is much better than for
nucleon decays in bound-unbound mirror pairs \cite{Tim05b}.

To check the validity of the approximation (\ref{rnres}) for
many-body systems
we have calculated ${\cal R}_{\Gamma}$ for bound-unbound
mirror states from Tables III and IV using the MCM  of  
the previous section. The width $\Gamma_{\alpha}$ have
been calculated by solving the Schrodinger-Bloch equation, as described in 
Ref. \cite{Des90}.
%The obtained solution for the residue of the R-matrix pole was corrected by
%the factor of xxx. This factor was negligible 
%for $^{19}$Ne($\frac{7}{2}^-_2$)
%state, about 1.01 for $^{19}$Ne($\frac{3}{2}^+_2$) and  noticably varied
%for $^{11}$C($\frac{3}{2}^-_3$), 1.01-1.11 and 1-1.08 for
%the $l_{\alpha}=0$ and $l_{\alpha}=2$ partial wave respectively.
The calculations have been done using two oscillator radii for potential  V2
and only one oscillator radius, 1.36 fm, for potential MN, because the larger
radius, $b$ = 1.6 fm, has caused numerical problems.
The resulting ratio ${\cal R}_{\Gamma}^{MCM}$ is presented in Table IV.
For $^{11}$B($\frac{3}{2}^+_3$)-$^{11}$C($\frac{3}{2}^+_3$) 
with $l$ = 2 and for 
$^{19}$F($\frac{3}{2}^+_2$)-$^{19}$Ne($\frac{3}{2}^+_2$)
${\cal R}_{\Gamma}^{MCM}$ agrees well both with 
 ${\cal R}_{PM}^{res}$ and ${\cal R}_{0}^{res}$. In the case of
  $^{11}$B($\frac{3}{2}^+_3$)-$^{11}$C($\frac{3}{2}^+_3$) 
with $l_{\alpha}$ = 0 ${\cal R}_{\Gamma}^{MCM}$ agrees only with
 ${\cal R}_{PM}^{res}$, deviating from  ${\cal R}_{PM}^{res}$ 
 by 12$\%$.
For $^{19}$F($\frac{7}{2}^-_2$)-$^{19}$Ne($\frac{7}{2}^-_2$),
a 68$\%$ difference between
${\cal R}_{\Gamma}^{MCM}$ and ${\cal R}_{0}^{res}$
is obtained. It originates because of the specific structure of
the second $\frac{7}{2}^-$ state in $^{19}$F ($^{19}$Ne) which
is mostly built on the second excited state $\frac{3}{2}^-$
of the $^{15}$N ($^{15}$O) core with an orbital
momentum $l=2$. The spectroscopic factor for the
configuration $\la ^{19}$F$| ^{15}$O$_{g.s.}\otimes \alpha \ra$
is very small,  about 10$^{-3}$. The spectroscopic factor
of the mirror configuration, defined 
 using the concept of the bound state approximation for
the narrow resonance function, is also very small. 
In such weak components effects due to charge symmetry breaking 
could be large. When the $^{15}$N$(\frac{3}{2}^-)
\otimes \alpha $ ($^{15}$O$(\frac{3}{2}^-)
\otimes \alpha $) configuration in $^{19}$F ($^{19}$Ne) is neglected, 
the  MCM gives for the
$\frac{7}{2}^-_2$ state  ${\cal R}_{\Gamma}^{MCM}$ values
that are close both to ${\cal R}_{0}^{res}$ and ${\cal R}_{PM}^{res}$.
For example, with V2 and an oscillator radius of 1.6 fm 
${\cal R}_{\Gamma}^{MCM}$ = 8.24$\times$10$^{-84}$ MeV$\cdot$fm.

\section{Unbound mirror pairs}

The ideas of Secs. II and III about mirror summetry  can be immediately
applied to the widths of two mirror narrow resonances 2 and 1.
For the ratio 
\beq
{\cal R}_{\Gamma \Gamma} = \Gamma_{\alpha}(2)/\Gamma_{\alpha}(1)
\eeqn{rgg}
Eqs. (\ref{rna}) and (\ref{rnres}) can be generalised straightforwardly
to give
\beq
{\cal R}_{\Gamma \Gamma} \approx {\cal R}_{\Gamma \Gamma}^0 = 
\frac{k_1}{k_2} \left| \frac{F_l(k_2 R_N)}{F_l(k_1 R_N)}\right|^2,
\eeqn{rngg}
where $k_i = \sqrt{2\mu E_i}/\hbar$ and $E_i$ is the resonance
energy of the $i$-th $\alpha$-particle. 
%Here $a_m$ is the
%radius at which the function $\left| \frac{F_l(k_2 a)}{F_l(k_1 a)}\right|$
%reaches its maximum and the approximation (\ref{rngg}) is expected
%to be reasonably accurate if $a_m$ is located at the nuclear surface.

The idea that the widths of two mirror resonances are related has 
already been used many times to predict unknown widths for those
resonances  where the widths of their 
mirror analogs  are known. The relation between
mirror widths  is usually obtained from the relation
of the width $\Gamma_{\alpha}$
to the Coulomb barrier penetration factor $P_l(E,R_N)$
and the reduced width $\theta_{\alpha}^2$ \cite{Lan58}:
\beq
\Gamma_{\alpha} = \frac{2\hbar^2}{\mu R_N^2} \,\theta_{\alpha}^2 \,
 P_l(E,R_N),
\eeqn{ga}
where
\beq
P_l(E,R_N) = \frac{k R_N}{F_l^2(kR_N)+G_l^2(kR_N)},
\eeqn{p}
$G_l^2(kR_N)$ is the irregular Coulomb function,
and $R_N$ is located somewhere on the surface.
Assuming that the reduced widths $\theta_{\alpha}(1)$ and
$\theta_{\alpha}(2)$ for mirror resonances are equal one
obtains from Eqs.(\ref{rgg}), (\ref{p}) and (\ref{ga}) 
\beq
{\cal R}_{\Gamma \Gamma}  \approx {\cal R}_{\Gamma \Gamma}^{\theta} 
\equiv \frac{k_2}{k_1} \,
\frac{F_l^2(k_1R_N)+G_l^2(k_1R_N)}{F_l^2(k_2R_N)+G_l^2(k_2R_N)}.
\eeqn{rggold}
The Eqs. (\ref{rngg}) and (\ref{rggold}) are not identical and can not be
deduced one from another.

\begin{figure}[t]
%\centerline{\psfig{figure=../../alpha.gg.fig1.eps,width=0.40\textwidth} }
\centerline{\psfig{figure=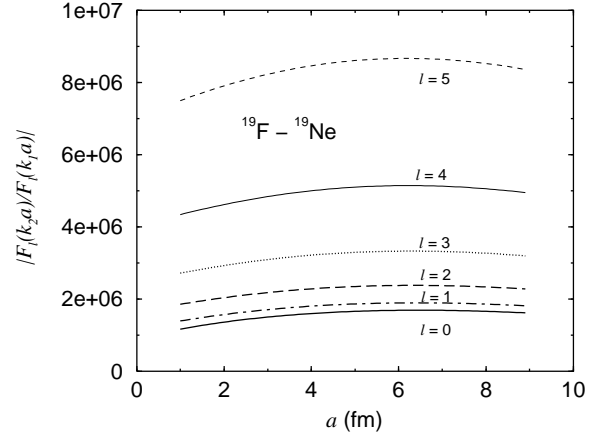,width=0.40\textwidth} }
\caption{ The ratio $ | F_l(k_2 a)/ F_l(k_1 a)| $  for 
$E_R(\alpha + ^{15}$N) = 0.350 MeV and 
$E_R(\alpha + ^{15}$O) = 0.850 MeV for different orbital momenta $l$
as a function of $a$.
}
\end{figure}

\begin{figure}[t]
%\centerline{\psfig{figure=../../rgg.n15.eps,width=0.40\textwidth} }
\centerline{\psfig{figure=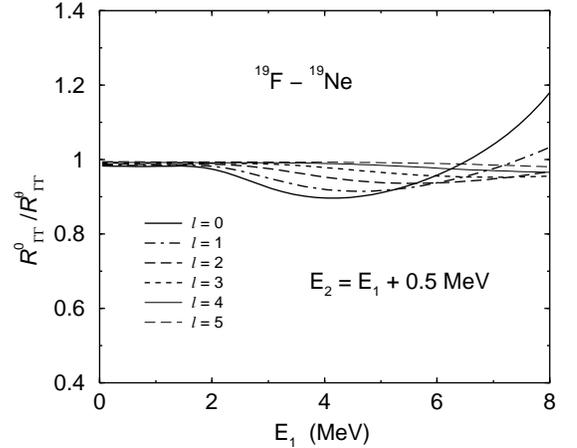,width=0.40\textwidth} }
\caption{ The ratio $ {\cal R}_{\Gamma \Gamma}^0/ 
{\cal R}_{\Gamma \Gamma}^{\theta}$  
 for different orbital momenta $l$ as a function
of the resonance energy    $E_1$ in $\alpha + ^{15}$N.
}
\end{figure}

First, we   investigate numerically the  difference between
the approximations (\ref{rngg}) and (\ref{rggold}) in a two-body
model for a hypothetical mirror pair $^{19}$F - $^{19}$Ne with 
arbitrary resonance energy $E_1$ in the $\alpha + ^{15}$N 
and ($E_2$) energy in the 
($\alpha + ^{15}$O) channel such that
$E_2 = E_1 + 0.5$ MeV,  for all $l_{\alpha} \leq 5$. 
The difference of about 0.5 MeV is typical for 
low-lying $\alpha$-particle
resonances in $^{19}$F - $^{19}$Ne. The 
 ratio $ | F_l(k_2 a)/ F_l(k_1 a)| $ for such a system is presented
in Fig. 5 for the lowest resonance energy in  the real 
$\alpha + ^{15}$N system, $E_1$ = 0.350 MeV, as a function of $a$.
This ratio is varies very slowly for $5 < a < 8$ fm and reaches
its maximum at about 6 - 7 fm, which is beyond the nuclear surface
radius $R_N$.
To compare (\ref{rngg}) and (\ref{rggold}) we calculate them
both at the surface, $R_N$ = 5 fm, as has been  done in other
studies of  mirror symmetry in the $^{19}$F - $^{19}$Ne resonances
\cite{Oli97,Dav03}.
The ratio ${\cal R}_{\Gamma \Gamma}^0/{\cal R}_{\Gamma \Gamma}^{\theta}$
is plotted in Fig.6 for different energies $E_1$ taken below the
Coulomb barrier. According to Fig.6, ${\cal R}_{\Gamma \Gamma}^0$
and ${\cal R}_{\Gamma \Gamma}^{\theta}$ are the same for $E_1 \leq 2$ MeV
but at higher energies a difference appears. This
difference increases with decreasing
orbital momentum. The largest difference, about 12$\%$, is seen
for $l_{\alpha} = 0$ at $E_1 \approx 4$ MeV. 
The most likely reason for this effect is
the growth of the resonance width with the resonance energy. 
At some point, the integral representation (\ref{gamma}) looses its
accuracy, making the approxiation (\ref{rngg}) invalid. The higher is the 
centifugal barrier, the higher the resonance energy can be before this happens.

\begin{table*}[t]
\caption{ 
Resonance widths $\Gamma_{\alpha}$ for mirror nuclei 1 and 2 (in MeV)
and their ratio 
calculated in the MCM, ${\cal R}_{\Gamma \Gamma}^{MCM}$,
and potential model, ${\cal R}_{\Gamma \Gamma}^{PM}$, 
for mirror states with 
spin-parity $J^{\pi}$ and orbital momentum $l$.
 } 
\begin {center}
\begin{tabular}{ p{0.8 cm} p{0.8 cm} p{3 cm} p{3 cm}
p{3 cm} p{2 cm} p{2 cm} p{1.6 cm} }
\hline 
 \hline
\\ 
& &  \multicolumn{3}{c}{{ Microscopic cluster model}} & 
 \multicolumn{3}{c}{{  Potential model}}\\
 \cline{3-8}
 $J^{\pi}$ & 
$l$  &  $\Gamma_{\alpha}(2)$ & 
 $\Gamma_{\alpha}(1)$ & ${\cal R}_{\Gamma\Gamma}^{MCM}$ &  
   $\Gamma_{\alpha}(2)$ & 
 $\Gamma_{\alpha}(1)$ & ${\cal R}_{\Gamma\Gamma}^{PM}$  \\
 \hline
\\
\multicolumn{8}{c}{ $^{7}{\rm Li}  - ^{7}{\rm Be}$ } \\
\\
$\frac{7}{2}^-_1$ &  3 &  0.142-0.267 & 0.079-0.149 & 1.795 $\pm$ 0.005 & 
0.247 & 0.134 & 1.82 \\ \\
\multicolumn{8}{c}
{ $^{11}{\rm B}  - ^{11}{\rm C}$ two-cluster MCM} \\ \\
  $\frac{5}{2}^-_2$ &2 & (1.68-4.21)$\times$10$^{-4}$ &
(1.07-2.56)$\times$10$^{-7}$ & 1610$\pm$40 & 6.47$\times$10$^{-3}$ &
4.51$\times$10$^{-6}$ & 1434 \\
 & 4 & (5.25-26.6)$\times$10$^{-7}$ & (5.28-26.6)$\times$10$^{-7}$ &
  (1.02$\pm$0.04)$\times$10$^4$ & 7.44$\times$10$^{-5}$ & 
 7.46$\times$10$^{-9}$ & 9964   \\
 $\frac{7}{2}^+_1$ & 3 & (2.19-7.20)$\times$10$^{-4}$ &   
 (5.78-18.5)$\times$10$^{-6}$ & 38.4 $\pm$ 0.5 & 6.19$\times$10$^{-3}$ &
 1.67$\times$10$^{-4}$ & 37\\
  & 5 & (0.82-8.19)$\times$10$^{-8}$ & (0.58-5.18)$\times$10$^{-10}$ &
 151 $\pm$ 7 & 5.38$\times$10$^{-5}$ & 3.54$\times$10$^{-7}$ & 152\\ \\
 \multicolumn{8}{c}
 { $^{11}{\rm B}  - ^{11}{\rm C}$ three-cluster MCM} \\
 \\
  $\frac{5}{2}^-_2$ &2 & 2.70$\times$10$^{-4}$ &
1.55$\times$10$^{-7}$ & 1740 &   &  &   \\
 & 4 & 1.24$\times$10$^{-6}$ & 1.08$\times$10$^{-10}$ &
  1.14$\times$10$^4$ &   &   &    \\
 $\frac{7}{2}^+_1$ & 3 & 1.60$\times$10$^{-3}$ &   
3.95$\times$10$^{-5}$ & 40.3 &   &   &  \\
  & 5 & 2.11$\times$10$^{-6}$ & 1.15$\times$10$^{-8}$ &
 183 &   &   & \\
 \\
 \multicolumn{8}{c}{ 
$^{19}{\rm F}  - ^{19}{\rm Ne}$} \\
 $\frac{7}{2}^+_1$ &  3 & (0.45-1.95)$\times$10$^{-8}$ &
   (0.36-1.50)$\times$10$^{-13}$ & 
  (1.28$\pm$0.03)$\times$10$^5$ & 1.23$\times$10$^{-6}$ & 
  9.50$\times$10$^{-12}$ & 1.29$\times$10$^5$   \\
  $\frac{5}{2}^-_2$ &  2 & (0.89-283)$\times$10$^{-7}$ &
 (0.48-134)$\times$10$^{-9}$ & 204$\pm$7 & 2.84$\times$10$^{-4}$ & 
 1.40$\times$10$^{-6}$ & 203 
 \\
\hline \hline
\end{tabular}
\end{center}
\label{table5}
\end{table*}

Next, we compare ${\cal R}_{\Gamma \Gamma}^0$
and ${\cal R}_{\Gamma \Gamma}^{\theta}$ to the results of
potential model and MCM calculations for some 
realistic mirror narrow resonances in $^{7}$Li - $^{7}$Be,
$^{11}$B - $^{11}$C and $^{19}$F - $^{19}$Ne.  Unlike in previous
sections, only one value of the diffuseness, 0.65 fm, has been used in
the potential model calculations. As for the MCM, the conditions
of the calculations are the same as in previous sections.

\begin{table*}[t]
\caption{ 
Analytical estimates  ${\cal R}_{\Gamma \Gamma}^{\theta}$ and
${\cal R}_{\Gamma \Gamma}^0$, MCM ratio
${\cal R}_{\Gamma \Gamma}^{MCM}$, potential model prediction
${\cal R}_{\Gamma \Gamma}^{PM}$ and the ratio 
${\cal R}_{\Gamma\Gamma}^{exp}$ of experimentally
known widths of mirror states in
$^7$Li-$^7$Be, $^{11}$B-$^{11}$C and $^{19}$F-$^{19}$Ne
with spin-parity $J^{\pi}$ and
orbital momentum $l$.
 } 
\begin {center}
\begin{tabular}{ p{2.0 cm} p{1 cm} p{1.2 cm} p{1.2 cm}
p{0.8 cm} p{1.5 cm} p{1.5 cm} p{2.8 cm}  p{2 cm}  p{2 cm}}
\hline 
\hline
\\ 
Mirror pair  &  $J^{\pi}$ &  $E_1$ & $E_2$ &
$l$  
&  ${\cal R}_{\Gamma \Gamma}^{\theta}$ & 
 ${\cal R}_{\Gamma \Gamma}^0$  & 
 ${\cal R}_{\Gamma\Gamma}^{MCM}$ &  
 ${\cal R}^{PM}_{\Gamma\Gamma}$ 
 & ${\cal R}_{\Gamma\Gamma}^{exp}$ \\
 \hline
 $^{7}{\rm Li}  - ^{7}{\rm Be}$ & $\frac{7}{2}^-_1$ & 
 2.1622 & 2.983 & 3 & 1.74 & 1.79 & 1.795 $\pm$ 0.005  & 1.82 & 1.88$\pm$0.24 \\
$^{11}{\rm B}  - ^{11}{\rm C}$ & $\frac{5}{2}^-_2$ & 0.2556 & 0.876 &
2 & 1493 & 1520 & 1660$\pm$80 & 1434 & 2140$\pm$970\\
 & & & & 4 & 9982 & 1.0$\times$10$^4$ & (1.06$\pm$0.08)$\times$10$^4$ &
 9964  & \\
  & $\frac{7}{2}^+_2$ & 0.5204 & 1.111 &
3 & 38.1 & 38.3 & 39.1$\pm$1.2 & 37.0 &  \\
 & & & & 5 & 152.3 & 152.2 & 163$\pm$20 & 151.8 & \\
$^{19}{\rm F}  - ^{19}{\rm Ne}$ & $\frac{7}{2}^+_1$ & 0.364 & 0.850 & 3 &
1.31$\times$10$^5$ & 1.30$\times$10$^5$ & (1.28$\pm$0.03)$\times$10$^5$ &
1.29$\times$10$^5$ & \\
 & $\frac{5}{2}^-_2$ &  0.6692 & 1.1826 & 2 & 209 & 207 & 204$\pm$7 & 203 &
 121$\pm$55\\
\hline \hline
\end{tabular}
\end{center}
\label{table6}
\end{table*}

The calculated widths $\Gamma_{\alpha}$
in mirror resonances and their ratio
are presented in Table V. In Table VI these ratios
are compared to ${\cal R}_{\Gamma \Gamma}^0$
and ${\cal R}_{\Gamma \Gamma}^{\theta}$.
In all cases studied, $\Gamma_{\alpha}$ depends strongly on the
choice of the model and its parameters.
For the $^7$Li-$^7$Be and $^{19}$F-$^{19}$Ne mirror pairs, 
the ratios  ${\cal R}_{\Gamma \Gamma}^{MCM}$ and 
${\cal R}_{\Gamma \Gamma}^{PM}$ agree  very well with the 
analytical predictions  ${\cal R}_{\Gamma \Gamma}^0$
and ${\cal R}_{\Gamma \Gamma}^{\theta}$.  For $^7$Li-$^7$Be they also
agree with experimental value ${\cal R}_{\Gamma \Gamma}^{exp}$ = 
$\Gamma_{\alpha}^{exp}(^7$Be$)/\Gamma_{\alpha}^{exp}(^7$Li) obtained
using the  $^7$Li and $^7$Be widths  of the $\frac{7}{2}^-$
resonance from \cite{Til02}. For the $\frac{5}{2}^-_2$ resonance
in $^{19}$F-$^{19}$Ne, the value  ${\cal R}_{\Gamma \Gamma}^{exp}$ 
= 121 $\pm$ 55
determined by using $\Gamma_{\alpha}^{exp}$ from 
\cite{Dav03}  is much smaller
than the theoretical values of 203 - 211. The most likely reason
for this is that the $^{19}$Ne($\frac{5}{2}^-_2$) width has been 
determined Ref. \cite{Dav03} indirectly
using the measured $^{19}$Ne($\frac{5}{2}^-_2$) branching
ratio $\Gamma_{\alpha}/\Gamma$ and its $\gamma$-width
 assuming that 
$\Gamma_{\gamma}(^{19}$F) = $\Gamma_{\gamma}(^{19}$Ne). Such an assumption
is not always valid.

For $^{11}$B-$^{11}$C, 
${\cal R}_{\Gamma \Gamma}^{PM}$ agrees very well with
the analytical predictions  ${\cal R}_{\Gamma \Gamma}^0$ and
${\cal R}_{\Gamma \Gamma}^{\theta}$. The two-cluster
MCM predictions also agree with them, expect for the $\frac{5}{2}^-_2$
state with $l_{\alpha}$ = 2 where a 10$\%$ increase in the ratio
of mirror widths can be seen. The three-cluster MCM increases this
ratio which could be due to the $^8$Be+t and
$^8$Be+$^3$He clustering effects.  Both the two- and three-cluster
predictions agree with the ratio of experimentally determined
widths taken from \cite{Ajz90}. In all cases, the
difference between the microscopic calculations and the analytical 
approximations (\ref{rngg}) and (\ref{rggold}) does not
exceed 10$\%$.

\section{Summary and conclusion}

In this paper, we have shown that the structureless
two-body  bound mirror systems $\alpha + X_1$
and $\alpha + X_2$, with the same strong  nuclear attraction
but different Coulomb repulsion, 
should have   ANCs that are related by a model-independent
analytical approximation (\ref{rna}). This expression involves the ratio of the
regular Coulomb wave functions calculated at imaginary momentum at
some distance $a$ between $\alpha$ and $X$. We have demonstrated that if
this distance is taken at the point where the product of $\alpha-X$ potential
and $\alpha-X$ wave function is the largest, which occurs around
$R_N \approx (1.1 - 1.3)(4^{1/3}+X^{1/3})$, then 
 deviation from this approximation should be small provided the
nuclear wave functions of these mirror systems are   similar
to each other  in the region that gives most contribution
to the ANC in Eq. (\ref{anc}). 
The analytical approximation (\ref{rna}) remains valid  for  
mirror systems with a many-body internal structure if mirror
spectroscopic factors are approximately the same and if
$X_1$ and $X_2$ 
are not too strongly deformed and/or do not have easily excited
low-lying states. 

The isospin symmetry between   mirror $\alpha$-decays
extends to bound-unbound and unbound mirror pairs.
In the first case, a link between the $\alpha$-particle ANC
of a bound state and the width of its mirror unbound analog
is given by the formula (\ref{rnres}). In the second case,
the link between the widths of mirror resonances can be given
by a new formula 
 (\ref{rngg}) that at the energies well below the combined Coulomb and
 centrifugal barrier
complements the old formula (\ref{rggold})
obtained using the concept of the  penetrability of the Coulomb barrier and
assuming  equality of the reduced widths of mirror
resonances. 

The comparison of the approximations (\ref{rna}), (\ref{rnres})
and (\ref{rngg}) to the results of exact calculations either in a
two-body potential model or in a microscopic cluster model 
for three mirror pairs, $^7$Li - $^7$Be, $^{11}$B - $^{11}$C
and $^{19}$F - $^{19}$Ne, have confirmed their validity for many
mirror nuclear states. The deviations from these approximations
are smaller than those seen in mirror nucleon decays in Ref.
\cite{Tim05a,Tim05b} because the difference in mirror $\alpha$-particle
wave functions are much smaller than the differences in mirror
proton and neutron wave functions, especially 
for loosely-bound states. The largest deviations from analytical
estimates have been seen for three-cluster $^{11}$B  - $^{11}$C
mirror states with excited $^7$Li and $^7$Be cores. Also,
a noticeable deviation has been seen for the second $\frac{7}{2}^-$
state in $^{19}$F-$^{19}$Ne. This state has  tiny spectroscopic
factors for the decay channels $\alpha + ^{15}$N$_{g.s.}$  and
$\alpha + ^{15}$O$_{g.s.}$ (about 0.001) and 
the probability of symmetry breaking in such week components
is always large.

The ANCs and $\alpha$-widths calculated in our microscopic approach are
sensitive to the model assumptions. In particular, they change within
a factor of four for different choices of the effective NN potential and
oscillator parameters, the smallest values being produced by combiningthe MN
potential with the oscillator parameter $b$ = 1.36 fm and the largest
values predicted by V2 with $b$ = 1.6 fm. The variation of ANCs and 
$\alpha$-widths with model assumptions can be even stronger if mirror states
have  specific structure, for example, the 
t+$^8$Be and $^3$He+$^8$Be configurations in 
$^{11}$B and $^{11}$C.
However, the calculated in the MCM ratios
${\cal R}$, ${\cal R}_{\Gamma}$ and  ${\cal R}_{\Gamma \Gamma}$
do not change much with different choices of unput model parameters.
This fact can be used to predict unknown ANCs or $\alpha$-widths
if the corresponding mirror quantities have been measured. 
Such predictions can be beneficial for nuclear astrophysics. 
Many low-energy $(\alpha,\gamma)$, $(\alpha,N)$ and $(N,\alpha)$
reactions proceed via the population of  isolated $\alpha$-particle
narrow resonances  the widths of which determine the corresponding
reaction rates. It is not always possible to measure such widths
because of the very small reaction cross sections involved. In this case,
using isospin symmetry in mirror $\alpha$-decays may be helpful.
For unbound mirror states this symmetry has already been  used.
For another class of mirror pairs, when 
the mirror analogs of the resonances are bound, 
$\alpha$-widths can be determined by measuring
the  ANCs of bound states
in $\alpha$-transfer reactions and using the relation
$\Gamma_{\alpha} =  {\cal R}_{\Gamma} C^2_{\alpha}$.
As an example,   we can point out that the widths of the astrophysically
important resonance $^{19}$Ne($\frac{3}{2}^+_2)$ at 4.033 MeV could
be detemined if the ANC of its mirror analog in $^{19}$F was known.
Unfortunately, available data on the $^{15}$N($^6$Li,d)$^{19}$F$^*(
\frac{3}{2}_2^+)$ reaction do not allow the extraction the ANC of interest
because of strong sensitivity to optical potentials and to the geometry
of the bound state potential well that arises due to angular 
momentum mismatch. An alternative possibility to measure this ANC 
with a high precision is to use the reaction
$^{15}$N($^{19}$F,$^{15}$N)$^{19}$F$^*$. This reaction involves the
same optical potentials in the entrance and exit channels and would
not suffer the angular momentum mismatch.

\section*{Acknowledgements}
%N.K.T. is grateful to Professors R.C. Johnson and I.J. Thompson for
%fruitful discussions and useful comments concerning this manuscript. 
Support from the UK EPSRC via grant GR/T28577 is gratefully acknowledged.

\section{Appendix} 

We prove here that $BW_2/a\phi_l^{(1)}$ is small with respect to
$A_0(a)$.
The coefficients $A$ and $B$ that are found from the continuity of 
$ \tilde{\phi}_l^{(2)}(r)$ and its derivative at $r = a$ can alternatively
be presented as follows:
\begin{eqnarray}
A = \frac{(W_2/a)'\phi_l^{(2)}- (W_2/a)\phi_l'^{(2)})}
{ (W_2/a)'\phi_l^{(1)}- 
(W_2/a)\phi_l'^{(1)}},
\label{A}\\
B = - \frac{\phi_l'^{(1)}\phi_l^{(2)} - \phi_l^{(1)} \phi_l'^{(2)}}{
\phi_l'^{(1)}(W_2/a) - \phi_l^{(1)} (W_2/a)'},
\label{B}\end{eqnarray}
where  $^{\prime}$ means 
differentiation with respect to $a$.
When expressed in terms of $F_1,\,F_2,\,$ and $W_2$ we find
\begin{eqnarray}
B = -\frac{\exp (\imath 
\delta_2)}{\kappa_2}\frac{F_2F_1^{\prime}-F_2^{\prime}F_1}
{W_2F_1^{\prime}-W_2^{\prime}F_1},
\label{B1}\end{eqnarray}
where $\delta _2 =-(l+1+\imath \eta_2)\pi /2$.
Therefore the quantity $BW_2/(a\phi_l^{(1)}A_0(a))$ is 
\begin{eqnarray}BW_2/(a\phi_l{(1)}A_0(a))=-\frac{F_1^{\prime}/F_1-F_2^{\prime}/F
_2}{F_1^{\prime}/F_1-W_2^{\prime}/W_2}.
\label{Baphi}\end{eqnarray}

We can get a good idea about the magnitude of this term by using semiclassical 
expressions for the $F_i$ and $W_2$.
For our purposes we can write
\begin{eqnarray}
W_2(a)=\frac{W_2(b)\exp(-\int_b^adr\,p_2(r))}{\sqrt{p_2(a)/p_2(b)}}, 
\label{W2cl}\\
F_i(a)=\frac{F_i(b)\exp(+\int_b^adr\,p_i(r))}{\sqrt{p_i(a)/p_i(b)}},
\label{Ficl}\end{eqnarray}
where the local wave numbers $p_i(r)$ are given by
\begin{eqnarray}
p_i(r)=\sqrt{\frac{2\eta_i \kappa_i}{r}+\frac{l(l+1)}{r^2}+\kappa_i^2},
\label{pi}\end{eqnarray}
and $b$ is an arbitrary point in the region where the semiclassical 
approximation is valid. We also assume that $a$ and $b$ lie in the region where 
the exponentially decreasing components of the $F_i$ can be ignored. 

Using these expressions and evaluating the derivatives in a way which 
consistently respects the semiclassical approximation (see \cite{Brink85}, pages 
23-24) we find
\begin{eqnarray}
BW_2/(a\phi_l^{(1)}A_0(a))=\frac{p_2(a)-p_1(a)}{p_2(a)+p_1(a)}.
\label{Baphic}\end{eqnarray}
For values of $a$ in the nuclear surface the difference $p_2(a)-p_1(a)$ tends to 
be very small fraction of $p_2(a)+p_1(a)$.
%For the $\alpha$-triton and $\alpha-^3$He pair I find $\mid 
%(p_2(a)-p_1(a))/(p_2(a)+p_1(a))\mid =0.004 $ at $a=3$fm and $\mid 
%(p_2(a)-p_1(a))/(p_2(a)+p_1(a))\mid =0.02 $ at $a=5$fm.   
Note that the condition $p_1(a)-p_2(a)=0$ is exactly the condition (in the 
semi-classical approximation) that $A_0(a)$ be a stationary function of $a$.

\end{document}